\documentclass{aa}  

\usepackage{graphicx}
\usepackage{txfonts}
\usepackage[]{hyperref}
\begin{document}

   \title{Improving the open cluster census.}

   \subtitle{II. An all-sky cluster catalogue with \emph{Gaia} DR3\thanks{Tables~\ref{tab:catalogue},~\ref{app:tab:all_crossmatches}, and the cluster members are only available in electronic form at the CDS via anonymous ftp to cdsarc.cds.unistra.fr (130.79.128.5) or via https://cdsarc.cds.unistra.fr/cgi-bin/qcat?J/A+A/}}


   \author{Emily L. Hunt\inst{1}\fnmsep\thanks{Fellow of the International Max Planck Research School for Astronomy and Cosmic Physics at the University of Heidelberg (IMPRS-HD).}
           \and
           Sabine Reffert\inst{1}
          }

   \institute{Landessternwarte, Zentrum für Astronomie der Universität Heidelberg, Königstuhl 12, 69117 Heidelberg, Germany\\
              \email{ehunt@lsw.uni-heidelberg.de}
             }

   \date{Received 1\textsuperscript{st} March 2023; accepted 21\textsuperscript{st} March 2023}

 
  \abstract
   {Data from the \emph{Gaia} satellite are revolutionising our understanding of the Milky Way. With every new data release, there is a need to update the census of open clusters.}
   {We aim to conduct a blind, all-sky search for open clusters using 729~million sources from \emph{Gaia} DR3 down to magnitude $G\sim20$, creating a homogeneous catalogue of clusters including many new objects.}
   {We used the Hierarchical Density-Based Spatial Clustering of Applications with Noise (HDBSCAN) algorithm to recover clusters. We validated our clusters using a statistical density test and a Bayesian convolutional neural network for colour-magnitude diagram classification. We inferred basic astrometric parameters, ages, extinctions, and distances for the clusters in the catalogue.}
   {We recovered 7167 clusters, 2387 of which are candidate new objects and 4782 of which crossmatch to objects in the literature, including 134 globular clusters. A more stringent cut of our catalogue contains 4105 highly reliable clusters, 739 of which are new. Owing to the scope of our methodology, we are able to tentatively suggest that many of the clusters we are unable to detect may not be real, including 1152 clusters from the Milky Way Star Cluster (MWSC) catalogue that should have been detectable in \emph{Gaia} data. Our cluster membership lists include many new members and often include tidal tails. Our catalogue's distribution traces the galactic warp, the spiral arm structure, and the dust distribution of the Milky Way. While much of the content of our catalogue contains bound open and globular clusters, as many as a few thousand of our clusters are more compatible with unbound moving groups, which we will classify in an upcoming work.}
   {We have conducted the largest search for open clusters to date, producing a single homogeneous star cluster catalogue which we make available with this paper.}

   \keywords{
   open clusters and associations: general 
   -- Methods: data analysis 
   -- Catalogs
   -- Astrometry
   }

   \maketitle
%
\section{Introduction}  

The Milky Way galaxy is an intricate ecosystem of ongoing star formation, evolution, and destruction. Open clusters (OCs) are one such part of this system, which form when molecular clouds condense into stars and may further condense into gravitationally bound groups of a few dozen to a few thousand stars. Hence, OCs offer an important way to study the immediate aftermath of star formation, as well as the ongoing evolution of stars up to an age of around $\sim$~1~Gyr, after which most OCs will have been broken up, with their member stars dissolving back into the galactic disk \citep{portegies_zwart_young_2010, krumholz_star_2019, krause_physics_2020}.

Our view of OCs has always been complicated by their sparsity and their typical location in the galactic disk, making them challenging to isolate from field stars along the line of sight \citep{cantat-gaudin_milky_2022}. However, dramatically improved astrometric and photometric data from the \emph{Gaia} satellite \citep{gaia_collaboration_gaia_2016} are revolutionising our understanding of OCs and the overall Milky Way. Compared with the \emph{Hipparcos} mission \citep{perryman_hipparcos_1997}, \emph{Gaia} provides order of magnitude improvements in proper motion and parallax accuracy for around $10^4$~times as many stars, with over 1 billion sources in total.

Because of these improvements, \emph{Gaia} has enabled many new insights into all properties of OCs. Works such as \cite{meingast_extended_2021} and \cite{tarricq_structural_2022} have shown that many nearby OCs have tidal tails or comas of ejected member stars indicative of their ongoing tidal disruption by the Milky Way. Other works such as \cite{bossini_age_2019} and \cite{cantat-gaudin_painting_2020} have used \emph{Gaia} photometry to infer cluster ages, extinctions, and distances, which can then be used to make wider inferences about the Milky Way, such as in \cite{castro-ginard_milky_2021} who used OCs to trace the spiral arms of the galaxy. Cleaned \emph{Gaia} cluster membership lists also improve spectroscopic studies such as \cite{baratella_gaia-eso_2020}, who combined \emph{Gaia} data with ground-based spectroscopic measurements to study the chemistry of OCs.

At the heart of all science with OCs, however, is the census of OCs itself. Particularly in the four years since \emph{Gaia} Data Release 2 \citep[DR2,][]{brown_gaia_2018}, many works have contributed major new insights into the census of OCs. Works such as \cite{cantat-gaudin_characterising_2018}, \cite{cantat-gaudin_clusters_2020}, and \cite{jaehnig_membership_2021} provide new membership lists for OCs with a significantly higher number of stars and reduced outliers from the field when compared to pre-\emph{Gaia} works. Thousands of new OCs have been reported using a range of unsupervised machine learning techniques, such as in \cite{castro-ginard_new_2018, castro-ginard_hunting_2019, castro-ginard_hunting_2020, castro-ginard_hunting_2022}, \cite{cantat-gaudin_gaia_2019}, or \cite{liu_catalog_2019}. The reliability of the census has also been improved, with works such as \cite{cantat-gaudin_clusters_2020} finding that a number of OCs discovered before \emph{Gaia} are likely to be asterisms.

One might wonder how much further \emph{Gaia} can improve the census of OCs, and what these improvements could reveal. In \cite{hunt_improving_2021} (hereafter Paper 1), we compare three different approaches for recovering OCs in \emph{Gaia} DR2 data, and find that the HDBSCAN clustering algorithm \citep[Hierarchical Density-Based Spatial Clustering of Applications with Noise,][]{hutchison_hdbscan_2013} is the most sensitive approach, although it is essential to reduce false positives with additional post-processing. In this work, we conduct the largest blind search for star clusters to date in \emph{Gaia} data, using \emph{Gaia} DR3 \citep{gaia_collaboration_gaia_2021}, methods developed in Paper 1, and additional validation criteria based on the photometry of every detected cluster. 

In Sect.~\ref{sec:data}, we describe the \emph{Gaia} DR3 data used in this work and the quality cuts we adopted to filter out unreliable sources. In Sect.~\ref{sec:clustering}, we briefly recap our clustering method from Paper 1 and tweaks made to improve cluster recovery within 1~kpc. We then outline a method to validate cluster candidates using their photometry in Sect.~\ref{sec:cmd_classifier}, which we generalise to additionally infer ages, extinctions, and photometric distances to our clusters in Sect.~\ref{sec:agenn}. In Sect.~\ref{sec:crossmatching}, we crossmatch our catalogue against literature works. Section~\ref{sec:results-overall} presents an overview of our catalogue. We discuss the non-detections of some literature clusters in Sect.~\ref{sec:discussion-undetected}, and discuss required steps a future work will take to improve the reliability of our new cluster candidates in Sect.~\ref{sec:discussion-moving_groups}. Section~\ref{sec:conclusion} summarises this work. 

During the preparation of this work, we found that many of the star clusters we detect appear much more compatible with unbound moving groups than bound OCs, regardless of the quality of their photometry or how strong of an overdensity they are. In an upcoming third paper, we will classify the clusters resulting from this work into bound and unbound clusters, which will result in our final catalogue. This work will follow shortly (Hunt~\&~Reffert,~\emph{in~prep.}).


\section{Data}\label{sec:data}  

In this section, we present a brief overview of \emph{Gaia} DR3 data and the preprocessing steps applied to prepare it for clustering analysis.

\subsection{\emph{Gaia} DR3}

The latest release of \emph{Gaia} \citep{gaia_collaboration_gaia_2016} astrometry and photometry, \emph{Gaia} DR3, presents an update to \emph{Gaia} DR2, based on an extra 12~months of data and various improvements to data processing. Astrometric and photometric data were released early in \emph{Gaia} EDR3 \citep{gaia_collaboration_gaia_2021}, with the full DR3 release containing other data products such as low-resolution spectra and updated radial velocities that we also make limited use of in this work \cite{gaia_collaboration_gaia_2022}. In total, DR3 contains 1.47~billion sources with 5- or 6-parameter astrometry, with a 30\% improvement in parallax precisions and a roughly doubled accuracy in proper motions. These improvements have a large impact on the detectability of OCs in \emph{Gaia} -- particularly for proper motions, where distant OCs have a signal-to-noise ratio (S/N) increased by a factor of $\sim4$ in \emph{Gaia} DR3 proper motion diagrams, owing to the halving in size of the Gaussian distribution of stars in both axes for distant clusters with proper motion dispersions smaller than \emph{Gaia} errors.

In addition, many improvements have been made to the processing and understanding of \emph{Gaia} data and systematics for \emph{Gaia} DR3. Most notably for OCs, \cite{lindegren_gaia_2021} provide a recipe for greatly reducing remaining parallax systematics for most sources in \emph{Gaia} DR3 down to a few $\mu$as in the best cases, which should significantly improve the accuracy of distances to the most distant clusters. \cite{cantat-gaudin_characterizing_2021} provide a recipe for correcting the proper motions of certain bright stars around $G\sim13$. While both of these corrections are too small to make a difference in unsupervised cluster searches, they are included in later cluster parameter determinations to improve the accuracy of final catalogue values.


\subsection{Outlier removal}\label{sec:data:outliers}

\begin{figure}[t]
   \centering
   \includegraphics[width=\columnwidth]{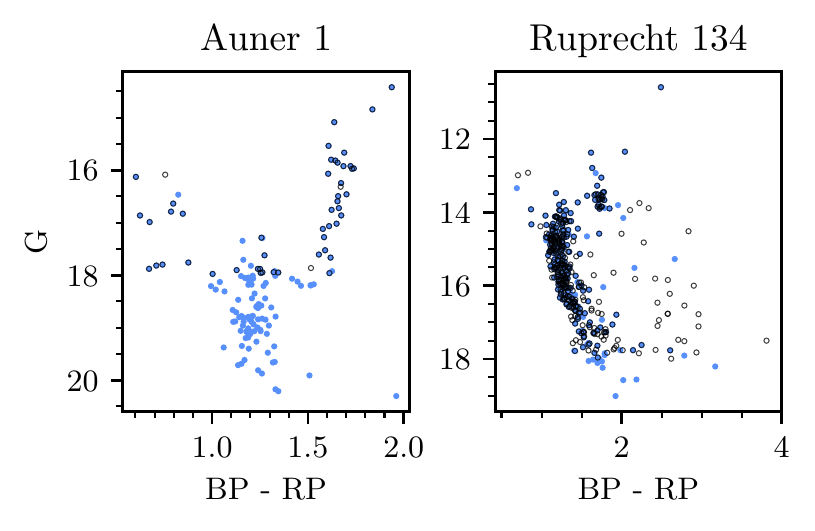}
   \caption{Comparison of cluster membership lists detected using \emph{Gaia} DR3 data cut at $G<18$ (black empty circles) and a \cite{rybizki_classifier_2022}~v1 criterion greater than 0.5 (blue filled circles) using separate runs of HDBSCAN and our pipeline for each cut, shown for Auner~1 (left) and Ruprecht~134 (right).}%
   \label{fig:cut_comparison}
\end{figure}

Despite improvements between \emph{Gaia} DR2 and DR3, many sources in the catalogue are still unreliable due to a number of reasons. For instance, blending in crowded fields can cause both astrometric and photometric errors, with sources being erroneously combined or split for any or all \emph{Gaia} measurements of the source. This is a particular issue in regions of the galactic disk with high numbers of sources. In addition, resolved and unresolved binary stars in DR3 may contribute significant errors to derived astrometric measurements for these sources, especially when their period is close to the one~year baseline used to measure parallaxes \citep{penoyre_astrometric_2022, lindegren_gaia_2021-1}, as well as causing issues with photometric measurements due to blending \citep{riello_gaia_2021, golovin_fifth_2023}.

To remove unreliable sources, a number of different quality cuts were investigated, both in isolation and combined: firstly, simple magnitude cuts, including $G<18$ as adopted in works such as Paper 1 and \cite{cantat-gaudin_characterising_2018}, $G<19$, and $G<20$; secondly, a cut on renormalised unit weight error (RUWE) values in the main \emph{Gaia} source table; and finally, a cut presented in \cite{rybizki_classifier_2022}, which uses a neural network and 17 diagnostic columns in the \emph{Gaia} EDR3 data release to classify astrometric solutions as reliable and unreliable, where we required a quality value of at least 0.5.

To evaluate the performance of these cuts, the reliability of cluster recovery with HDBSCAN \citep[Hierarchical Density-Based Spatial Clustering of Applications with Noise, ][]{hutchison_hdbscan_2013, mcinnes_hdbscan_2017} was inspected manually for 15 challenging to detect clusters given different combinations of these cuts. Notable clusters in this process include Ruprecht~134, a difficult to recover cluster located in the most crowded region of the galactic disk at $l,b=(0.28^\circ,-1.63^\circ)$ and at a distance of $\sim3$~kpc, in addition to a number of clusters reported in \cite{cantat-gaudin_clusters_2020} but not detected in Paper 1 in \emph{Gaia} DR2, such as Berkeley~91 and Auner~1.

A single, magnitude-independent cut based only on the quality flag of \cite{rybizki_classifier_2022} was found to outperform all other cuts trialed for cluster recovery. On average, for the trial set of 15 clusters, clusters recovered using this cut had the highest S/N of any recovered by any of the trialed cuts, with S/Ns being an average of 65\% higher than clusters recovered using the $G<18$ cut common in the literature \citep[see e.g.][]{cantat-gaudin_characterising_2018, castro-ginard_hunting_2022}. Clusters almost always had more member stars than a simple $G<18$ cut, with up to around twice as many member stars for distant, faint clusters where only giant stars can be resolved for magnitudes $G<18$, such as for the distant cluster Auner~1 at a distance of 6.8~kpc. Inevitably, this cut should result in more complete membership lists and a more complete overall catalogue of clusters. 

As a visual example, the CMDs of Auner~1 and Ruprecht~134 from clustering analyses using this cut and a $G<18$ cut are compared in Fig.~\ref{fig:cut_comparison}. Auner~1 is a distant and difficult to detect cluster, for which only 51 stars are detected in the $G<18$ trial for a cluster S/N of 10.8$\sigma$. However, the Rybizki cut cluster includes many additional faint sources, for a total of 139 member stars and an improved S/N of 17.9$\sigma$. In the case of Ruprecht~134, a massive cluster in a crowded region near the galactic centre, the Rybizki cut cluster has fewer sources than the $G<18$ cut (277 to 355) but a higher S/N (24.7$\sigma$ to 16.6$\sigma$), with the Rybizki cut removing a number of spurious sources from the cluster membership and the field -- improving the cluster membership list and the cluster's contrast against field stars.

Compared to having no cut at all, adoption of this cut typically has a minimal impact on the number of member stars for all clusters -- it appears that sources with unreliable astrometry are already so unreliable that their position in 5D \emph{Gaia} astrometry is too far from the bulk cluster position to be tagged as members, and few outliers are removed from cluster CMDs by this (or any) cut. Instead, in the crowded region at the galactic centre around Ruprecht~134, 85\% of the sources in this field were removed by the cut, yet all reliable clusters in this field \citep[including the nearby UFMG~88 reported by][]{ferreira_new_2021} remained with a similar membership list to with no cut at all. In addition, the lack of a magnitude cut means that in sparse fields where faint sources have reliable astrometry, clusters such as the high galactic latitude Blanco~1 have membership lists down to fainter than $G\sim20$, two magnitudes fainter than the membership list of \cite{cantat-gaudin_clusters_2020} for this cluster. 

Only the v1 version of the \cite{rybizki_classifier_2022} quality flag was available during preparation of cluster membership lists in this work, for which a minimum value of 0.5 was adopted. Later versions of the initial \cite{rybizki_classifier_2022} pre-print and eventual published paper have a slightly improved version of the quality flag, although in practice it was found to make a negligible difference to the final results of this work and so clustering analysis was not revised to include it.

In total, 729.7 million sources in \emph{Gaia} DR3 have a \cite{rybizki_classifier_2022} v1 quality flag of at least 0.5 and were selected for further clustering analysis in this work. This represents significantly more sources than the 301.7 million sources with $G<18$, a cut adopted in works such as \cite{castro-ginard_hunting_2022} or \cite{cantat-gaudin_clusters_2020}, and should result in a greater total number of both detected clusters and member stars.


\subsection{Data partitioning}

Finally, due to computational reasons, we partition the \emph{Gaia} dataset into three separate collections for further analysis, as it is not possible to efficiently perform clustering analysis with 729.7 million sources at once. We aim to divide the \emph{Gaia} dataset in such a way so that no more than 20 million sources are in any one field and so that a cluster of around $20 pc$ tidal radius can always be reliably detected regardless of its distance or location within adopted fields, which should be a reasonable upper size limit for almost all OCs based on \cite{kharchenko_global_2013} and \cite{cantat-gaudin_clusters_2020}.

As in Paper 1, the HEALPix (Hierarchical Equal Area isoLatitude Pixelation) tessellation scheme was used to segment the entire \emph{Gaia} dataset \citep{gorski_healpix:_2005}, with calculations performed by the Python package \texttt{Healpy} \citep{zonca_healpy_2019}. This has advantages over other methods to subdivide spheres into a finite number of regions, in that all regions at a given tessellation level have the same area, and spherical distortions are minimised. However, unlike in Paper 1, the origin of the HEALPix grid was set at the origin of galactic coordinates ($l,b=(0^\circ,0^\circ)$), instead of the default ICRS origin at right ascension and declination values of $\alpha,\delta=(0^\circ,0^\circ)$ used in \emph{Gaia} data releases, as this places most remaining spherical distortions at high galactic latitudes where we expect to find few clusters, meaning that all fields on the most important regions of the galactic disk are simple quadrilaterals.

We adopted three different partitioning schemes to detect clusters in three different distance ranges: those more distant than 750~pc, those closer than 750~pc, and those closer than 150~pc. Each scheme used large enough fields to detect clusters at each different distance range, but while minimising the number of stars in each field to keep the fields feasible to perform clustering analysis on. Firstly, for the most distant clusters, we adopted the same methodology as in Paper 1, dividing the entire \emph{Gaia} dataset into 12288 HEALPix level five pixels. To avoid losing clusters on the edge of each pixel, each pixel is grouped into fields containing the pixel itself and its eight nearest neighbours, effectively overlapping each $\approx5.5^\circ \times 5.5^\circ$ field by $1.8^\circ$ with all surrounding neighbours, with every pixel appearing in nine separate fields and in the centre of one. Next, to detect clusters closer than 750~pc, a HEALPix level two scheme with 192 pixels was adopted, containing only sources with $\varpi>1$~mas, using the same nine pixels per field system and resulting in overlapping fields of size~$\approx44^\circ \times 44^\circ$. Finally, for clusters closer than 150~pc, which can have large extents on the sky, a single field containing all stars closer than 250~pc was used, based on photo-geometric distances to sources in \cite{bailer-jones_estimating_2021}.

Between these three systems, all bound members of all open clusters of size $20 pc$ or smaller should be contained within these fields -- although in reality, this is only a worst-case constraint at the 750~pc and 150~pc crossover points and for a cluster in the worst possible location in a field, and many significantly larger clusters (including tidal tails many times their size) would be detectable in other regions.


\section{Cluster recovery}\label{sec:clustering}  

Next, we discuss the methodology we adopted to recover clusters in \emph{Gaia} data, assign basic parameters, and crossmatch to existing cluster catalogues in the literature.

\subsection{HDBSCAN}

Many different algorithms have been used to date to recover clusters in \emph{Gaia} data. We present a review and full explanation of these algorithms in Paper 1, in which we found that the HDBSCAN algorithm \citep{hutchison_hdbscan_2013, mcinnes_hdbscan_2017} is the most sensitive for recovering OCs in \emph{Gaia} data. 

Briefly, HDBSCAN is an updated version of the DBSCAN algorithm \citep{ester_density-based_1996}, for which only a minimum cluster size $m_{clSize}$ and minimum number of points in the neighbourhood of a cluster core point $m_{Pts}$ must be specified, unlike DBSCAN which instead uses $m_{Pts}$ and a minimum, global distance between points in a cluster $\epsilon$. DBSCAN has seen much use in the literature so far for OC recovery, such as in \cite{castro-ginard_new_2018, castro-ginard_hunting_2019, castro-ginard_hunting_2020, castro-ginard_hunting_2022} or \cite{he_catalogue_2021, he_new_2022}. The main challenge of DBSCAN is that $\epsilon$ must be set globally for an entire dataset, which can limit the sensitivity of the algorithm for datasets of varying density -- such as the \emph{Gaia} dataset, which has different densities at different distances and locations within the galaxy. 

Instead, HDBSCAN copes with varying density datasets by effectively considering all possible DBSCAN $\epsilon$ solutions for all regions of a dataset, selecting the best clusters based on the lower limit of cluster size $m_{clSize}$. HDBSCAN has so far been used to detect moving groups in \emph{Gaia} data by \cite{kounkel_untangling_2019} and \cite{kounkel_untangling_2020}, as well as being used to find 41 new OCs in Paper 1, and being used by \cite{tarricq_structural_2022} to reveal new tidal tails and comas of numerous OCs within 1.5~kpc. HDBSCAN has not yet been used to conduct a search through all \emph{Gaia} data for OCs.

A major flaw of HDBSCAN, however, is its high false positive rate. In Paper 1, we show that this is due to the algorithm being overconfident, reporting dense random fluctuations of a given dataset as clusters. To mitigate this, we adopt the cluster significance test (CST) from Paper 1, which searches for field stars surrounding a cluster and compares the nearest neighbour distribution of cluster stars with that of field stars. This then produces a signal-to-noise ratio (S/N), with CST scores greater than $5\sigma$ corresponding to highly likely clusters.

The issue of how to convert the five dimensions of \emph{Gaia} astrometry into a form best usable by a clustering algorithm is an open problem. Converting proper motions and parallaxes to velocities and distances respectively is one such approach \citep[e.g. as in][]{kounkel_untangling_2020, he_new_2022}, although a major issue is that converting \emph{Gaia} parallaxes to distances is non-trivial and results in asymmetric errors and non-Gaussian parameter distributions \citep{bailer-jones_estimating_2021}. Instead, we use the approach adopted in Paper 1, similar to that of works such as \cite{castro-ginard_new_2018} and \cite{liu_catalog_2019}. We use \emph{Gaia} positions, proper motions, and parallaxes directly, but with two preprocessing steps: firstly, recentring them into a coordinate frame with an origin at the centre of each respective field, which removes spherical distortions present at high declinations; secondly, rescaling all five axes of the dataset to have the same median and interquartile range, effectively removing the units of each axis of the data. Particularly for HDBSCAN, which can cope with varying density datasets, the choice to use these five simple recentred and rescaled features was found to have no impact on the detectability and membership lists of nearby clusters, while having great benefits for clusters more distant than $\sim2$~kpc, for which a distance-based approach causes many clusters to have sparser, non-Gaussian, and more challenging to detect distributions.

The one exception to this in this work is for the single field of all stars within $250$~pc, which was adopted to help improve the accuracy of cluster membership lists for very nearby clusters with large angular extents on the sky such as the Hyades. Given that this field covers the entire sky, it is not possible to avoid high latitude spherical distortions with a simple recentring; instead, photo-geometric distances from \cite{bailer-jones_estimating_2021} were used to convert positions and parallaxes to a Cartesian coordinate frame, with proper motions converted to tangential velocities. At such small distances, the uncertainties in \cite{bailer-jones_estimating_2021} are small and not prior-dominated, and so reliance on \emph{Gaia}-derived distances for the single nearby field should not cause any issues.



\subsection{Clustering analysis and catalogue merging}

\begin{figure}[t]
   \centering
   \includegraphics[width=\columnwidth]{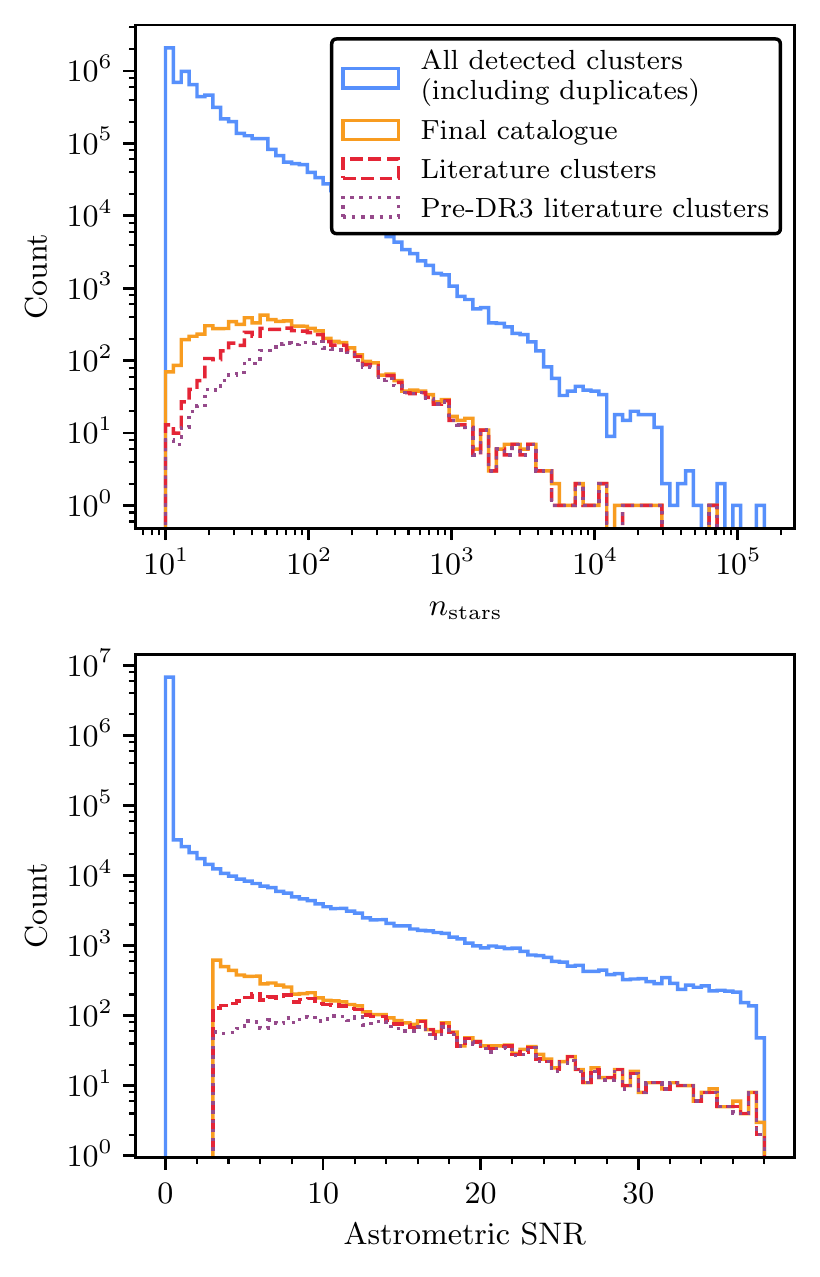}
   \caption{Statistics of all detected clusters compared against the final catalogue. \emph{Top}: distribution of the number of member stars of detected clusters, $n_\text{stars}$, for all detected clusters in all fields before catalogue merging and duplicate removal (solid blue line), for the final catalogue (solid orange line), and amongst clusters in the final catalogue that crossmatch to clusters in the literature, for all literature clusters (solid red line) and for only those detected before the release of \emph{Gaia} EDR3 (dotted purple line). \emph{Bottom:} as above, but for the astrometric S/N (CST score) for all clusters in these sets. S/Ns have a maximum value of 38 due to numerical reasons.}%
   \label{fig:cst_histogram}
\end{figure}

Using HDBSCAN and the same range of parameter choices as in Paper 1 ($m_{clSize}\,\epsilon\,\{10,20,40,80\}$, $m_{Pts}=10$), clustering analysis on all HEALPix level two and five fields was completed in around eight days of runtime on a machine with a 48 core Intel(R) Xeon(R) E5-2650 CPU with 48 GB of RAM. This run was mostly RAM-limited due to the worst-case $\mathcal{O}(n^3)$ memory use of the HDBSCAN implementation used for the largest fields. Given that fields overlap and that different parameter choices can detect the same cluster, each cluster can be duplicated up to four times within a single field, up to nine times by appearing in all neighbouring fields and a further time by appearing in different distance ranges (if the cluster has a distance between 0.7 to 1~kpc, or less than 250~pc). Hence, in the worst case, a single cluster could be duplicated 72 times. It is essential and non-trivial to merge the results of all fields accurately and without losing or duplicating any one individual cluster.

In total, 7.1~million different clusters were detected (including duplicates), almost all of which are astrometric false positives due to the oversensitivity flaws of HDBSCAN discussed in Paper 1. These clusters can be removed by using their astrometric S/N, as derived by the CST. Figure~\ref{fig:cst_histogram} shows histograms of the S/Ns of detected clusters, showing a clear spike in count for $\text{S/N}<0.5$ and an increasing trend in S/N for $\text{S/N} \lesssim 3$ that deviates from the relatively straight log-linear relation in S/N present for $\text{S/N}>3$, suggesting that an additional component of false positives is contributing to the otherwise log-linear component of reliable astrometric clusters at low S/Ns. This figure, our results from Paper~1, and the poor quality of the low-S/N clusters we detect strongly support that most low-S/N clusters are false positives; however, exactly where to set an S/N threshold is a non-trivial decision that has a large effect on the rest of the catalogue. A catalogue can choose to prioritise completeness, having a low threshold and including as many true positives as possible, but while inevitably including many false positives and sacrificing precision; or, a catalogue can do the opposite, having a lower completeness but also minimal false positives and maximised reliability of all objects in the catalogue.

For the purposes of this work, we chose to prioritise the precision and reliability of the catalogue, adopting a higher threshold on the minimum S/N of clusters. This sacrifices some completeness so that all final catalogue entries are likely to be real astrometric overdensities and not mere statistical fluctuations. This approach also comes with a key advantage. Our field tiling strategy aimed to prevent any real clusters from being `lost', aiming to recover $>99\%$ of real, good-quality OCs in a single catalogue. However, merging the results of so many separate clustering runs is a difficult and non-trivial task, and early experiments showed that the inclusion of false positives in the catalogue had a severe effect on the reliability and accuracy of the catalogue merging process. It was common that false positives and clear real OCs would share members in different clustering runs, meaning that low S/N thresholds on the final catalogue would adversely affect the catalogue's completeness at higher S/Ns. For the purposes of this work, we set a higher threshold on the minimum S/N, requiring $\text{S/N}>3\sigma$. This cut was found to maximise the quality of later catalogue merging steps, while removing a high number of false positives and retaining reliable clusters. Many false positives share member stars with real OCs, which greatly complicated the merging process and made the choice of which cluster to keep challenging. A single S/N cut means that our incompleteness is well characterised and easy to understand, whereas lower cuts were found to adversely affect catalogue completeness even at high S/Ns in a difficult to characterise way. In addition, while our adopted cut is at an S/N of $3\sigma$, clusters with an S/N lower than even $5\sigma$ may have minimal scientific usefulness, as they cannot be asserted as being real astrometric overdensities beyond any reasonable doubt; as such, it is not worth including such clusters in the catalogue at the expense of the recovery of better, real objects.

Inevitably, some low-S/N real OCs are likely to be lost in this process. We discuss the number of literature objects that are lost due to this cut in Sect.~\ref{sec:discussion-undetected:methodological-reasons:2-cuts}, and we briefly discuss some of the improvements to clustering algorithms that could be used to simplify the merging process and entirely remove the need for an S/N cut to ensure the catalogue's reliability in Sect.~\ref{sec:conclusion}.



After dropping unreliable low S/N clusters, the results of each parameter run in every field were merged. For clusters where every $m_{clSize}$ detected an identical object, duplicates were simply dropped. In some cases (such as for the largest OCs and GCs), smaller $m_{clSize}$ runs may split the cluster into two subclusters. Generally, it was possible to remove duplicate small subclusters by only keeping the single largest cluster. This process was extensively checked by hand, keeping smaller clusters instead in the case of some binary and coincident clusters which are better selected as being split, which was aided by fitting Gaussian mixture models to every cluster and evaluating the Bayesian information criterion of one and two-component fits, flagging clusters where a two component fit was preferred for potential splitting.

Secondly, cluster duplicates between fields must be removed. Using maximum likelihood distances calculated with the method presented in \cite{cantat-gaudin_characterising_2018}, clusters likely to be affected by edge effects or likely to be better detected at a different HEALPix level were removed. Clusters from the 250~pc run were only kept if they were closer than 175~pc. Clusters from the HEALPix level 2 run were only kept with distances between 150 and 750~pc. Finally, clusters from the HEALPix level 5 run were only kept if they had distances greater than 700~pc. The small overlaps in these distance ranges allow the best cluster to be selected later for clusters on the boundaries.

Next, duplicate clusters due to the overlap between fields must be removed. As each field is composed of nine pixels, a cluster can appear in up to nine separate fields. Keeping only clusters in the central pixel of every field is sufficient to mostly remove duplicates, retaining only the best cluster detection in the central pixel where edge effects are minimised. However, cluster membership lists are often not identical between fields, and it is hence possible that a cluster's mean position could be different enough between runs to appear in the central pixel of multiple fields or to never appear in the central pixel of any field. Particularly for small clusters of 20 stars or less, the inclusion or removal of even a single star can have a reasonable impact on the mean position of the cluster. This effect is worst for the nearest clusters with the largest angular extents on the sky relative to the field they are in. While this effect only impacts a small number of clusters (causing around $\sim1\%$ of clusters reported in \cite{cantat-gaudin_clusters_2020} to be lost), it is nevertheless important to address to ensure the final catalogue is as complete as possible.

To mitigate this effect, clusters near to the edge of a central pixel were also kept. After extensive testing, it was found that cluster positions generally vary by no more than $\sim 1$~pc at the distance of the cluster between different fields. We adopt a more tolerant cut corresponding to $\sim 5$~pc for a cluster at a worst-case distance, such that clusters within 1.91$^\circ$ (HEALPix level 2) or 0.41$^\circ$ (HEALPix level 5) of the edge of a central pixel were also kept. This is small compared to the overall field sizes of $\approx44^\circ \times 44^\circ$ (HEALPix level 2) or $\approx5.5^\circ \times 5.5^\circ$ (HEALPix level 5), but was nevertheless found to be sufficient to avoid losing any genuine clusters.

These processes removed most duplicated clusters while minimising the number of clusters lost during the merging process, although some duplicates still remained within the allowed overlaps between fields. These clusters were removed by looking for clusters with similar membership lists, mean positions, mean proper motions, and mean parallaxes, and selecting the cluster in each case with only the highest distance from any field edge. This process was also verified extensively by hand. For 23 large clusters (typically with tidal tails larger than the field they are in), duplicate clusters were similar but with both having additional members. In these cases, the clusters were merged into single clusters.

Finally, the catalogue was checked for clear, known binary clusters that were not correctly split by HDBSCAN. Four probable cases were identified, including the close binary Collinder~394/NGC~6716 as well as UBC~76/UBC~77. Generally, these binary clusters had very similar proper motion and parallax distributions, making them difficult or impossible for the HDBSCAN algorithm to split -- particularly since HDBSCAN cannot assign members to two clusters at once, although this is necessary for such close and difficult to separate objects. These clusters were split with Gaussian mixture models by selecting the number of components with the highest Bayesian information criterion. In all four cases, multiple components were preferred over a single component. It is likely that some other objects in the catalogue may also be better described as binary clusters, although this would need to be investigated carefully on a case-by-case basis \citep[see e.g.][]{kovaleva_collinder_2020, anders_ngc_2022-1} or with analysis using improved astrometry of a future \emph{Gaia} data release. This resulted in a list of 7788 clusters for further analysis.


\subsection{Additional parameters and membership determination}\label{sec:clustering:parameters}

Cluster parameters were mostly determined following the same approach as in Paper 1. However, it was noticed that many clusters are detected with tidal tails or comas, despite this study not being initially designed to detect cluster tidal tails. This is particularly common for clusters within $\sim2$~kpc. In many cases, this can cause clusters to have strongly biased mean parameters, such as for the cluster Mamajek~4 at a distance of 444~pc. Mamajek~4 has a tidal tail that stretches for 15$^\circ$ or 100~pc from its core, although only one side of the tail is detected due to limitations of the size of the field it was detected in. Using a simple mean position and proper motion for such clusters is hence affected by this asymmetry and is strongly biased. 

Instead, we aim to derive cluster parameters for the central part of clusters only. In practice, particularly for dissolving clusters with a majority of their mass in their tidal tails, it can be difficult to decide where stars should be called members of the cluster or members of the field. For instance, \cite{tarricq_structural_2022} attempted to derive structural parameters for 467 OCs within 1.5~kpc, but their method (based on fitting \cite{king_structure_1962} profiles) only succeeded on 389 clusters. To allow for accurate parameters to be inferred for all clusters homogeneously, we adopt a simple methodology comparing the density of cluster members with that of the field.

Firstly, cluster members with a HDBSCAN membership probability of less than 50\% were discarded. HDBSCAN membership probabilities are not based on \emph{Gaia} uncertainties, but rather only on the proximity of a given member to the bulk of the cluster. It was noticed that membership probabilities lower than this limit always correspond to low-quality cluster members or members of tidal tails, and are hence not worth including in the determination of reliable parameters of clusters. 

Next, using these members, cluster centres are derived in a way insensitive to asymmetries. Kernel density estimation was used to select the modal point of the cluster stellar distribution, with a bandwidth set to 1~pc at the distance of the cluster. 

Finally, using this cluster centre, the radius at which the overall cluster has the best contrast to field stars was selected. In practice, this is similar to the \cite{king_structure_1962} definition of tidal radius as the radius at which a cluster's density begins to exceed that of the density of the field, but is model-independent and can be easily and efficiently computed for the entire catalogue by selecting the radius at which a cluster has the highest CST against field stars. For instance, for well-defined clusters such as the Pleiades and Blanco~1, this radius was found to exclude cluster tidal tails while corresponding well with literature tidal radius values in \cite{kharchenko_global_2013} (see Sect.~\ref{sec:results-overall} for a discussion of our cluster radii.)

Mean parameters such as mean proper motion and parallax were then calculated given the members within the cluster's estimated tidal radius, in addition to maximum likelihood cluster distances calculated using the method of \cite{cantat-gaudin_characterising_2018}. To calculate more accurate distances, the parallax bias of member stars was corrected using the method in \cite{lindegren_gaia_2021}, which improved the accuracy of cluster distances particularly for distant clusters. As the \cite{lindegren_gaia_2021} parallax correction can only be applied for certain parameter ranges, for six clusters, too few sources (or no sources) had available corrections, and so we applied a simple global offset of $\varpi_0 = -17$~$\mu$as as derived in \cite{lindegren_gaia_2021}. These six clusters are flagged in the final catalogue as having less accurate distances. Overall, although the \cite{cantat-gaudin_characterising_2018} distance method assumes that the size of clusters is negligible compared to their distance, which introduces a bias for nearby clusters, our astrometric cluster distances were nevertheless found to agree well with the literature. For instance, we derive a distance of $47.19^{+0.004}_{-0.005}$~pc to the Hyades, which is comparable to the $47.34\pm0.21$~pc distance in \cite{mcarthur_astrometry_2011}, who use Hubble Space Telescope parallaxes to a subset of Hyades member stars to derive its distance.

In addition, \cite{king_structure_1962} core radii were estimated given our estimated tidal radius $r_{t}$ and radius containing 50\% of members of the core $r_{50}$, since there exists only one solution to the number density equation in \cite{king_structure_1962} (Eqn.~18) given $n(r_{50})$ and $r_{t}$. While approximate and less accurate than full Markov chain Monte-Carlo (MCMC) fits such as those performed in \cite{tarricq_structural_2022}, these core radii still provide a good approximation of a \cite{king_structure_1962} model fit and compared well to literature values for well-defined clusters for which different works have similar membership lists. Having calculated basic astrometric parameters for our clusters, we next calculate photometric parameters for our clusters using convolutional neural networks.


\section{Photometric validation}\label{sec:cmd_classifier}  

\begin{table}
\caption{Probability distributions used for simulated clusters for training of the CMD classifier.}
\centering
\label{tab:simulated_oc_distributions}
\begin{tabular}{c c c}
\hline\hline
Param. & Range & Distribution \\
\hline

$\log t$ & $[6.4,\,10.0]$ & $\mathcal{U}(6.4,\,10.0)$ \\
$[\text{Fe}/\text{H}]$ & $[-0.5,\,0.5]$ & $\mathcal{B}(4.0,\,4.0) - 0.5$ \\
$m - M$ & $[3.2,\,15.73]$ & $\mathcal{U}(3.2,\,15.73)$ \\
$A_V$ & $[0.0,\,8.0]$ & $\mathcal{B}(\sqrt{d/3},\,\sqrt{d/5}) \cdot 8 \tanh(d / 2)$\tablefootmark{a} \\
$n_{stars}$ & $[10,\,10000]$ & $10^{3 \cdot \mathcal{B}(2,\,3.5) + 1}$ \\
$\sigma_{\Delta A_V}$ & $[0.0,\,0.6]$ & $0.4 \cdot \mathcal{T}(1.25)$ \\
$l$ & $[0^\circ,\,360^\circ)$ & $\mathcal{U}(0,\,360)$ \\
$b$ & $[-90^\circ,\,90^\circ]$ & $90\cdot\mathcal{S}\cdot\mathcal{R}(\mathcal{B}(1,\,35),\,\mathcal{B}(1,\,12),\,2/3)$ \\

\hline

\end{tabular}

\tablefoot{Distributions of parameters are quoted as uniform distributions $\mathcal{U}(a,\,b)$ between $a$ and $b$, beta distributions $\mathcal{B}(a,\,b)$ with parameters $a$ and $b$, truncated exponential distributions $\mathcal{T}(a)$ truncated at $a$, $\mathcal{R}(a,\,b,\,x)$ which is a weighted choice with probability $x$ of choosing value $a$ and probability $1-x$ of choosing value $b$, and $\mathcal{S}$ which is a random sign with value $+1$ or $-1$. \tablefoottext{a}{Distances $d$ in kpc.}
}

\end{table}

In this section, we use photometry to validate members of the cluster catalogue as being compatible with single-population OCs and infer basic parameters, entirely using neural networks and simulated data. While \cite{castro-ginard_new_2018, castro-ginard_hunting_2019, castro-ginard_hunting_2020, castro-ginard_hunting_2022} successfully use neural networks to classify candidate clusters as real or false with their photometry, and while \cite{cantat-gaudin_painting_2020} and \cite{kounkel_untangling_2020} use neural networks to infer the ages, extinctions, and distances of their catalogued clusters, all of these works rely partially or entirely on existing examples of OCs detected in \emph{Gaia}. 

While such an approach mitigates issues with simulated training data, namely that stellar isochrones such as \cite{bressan_parsec_2012} are typically an imperfect fit to the observed CMDs of OCs \citep{cantat-gaudin_painting_2020}, it is difficult to guarantee that a small training dataset that relies mostly or entirely on examples of OCs from \emph{Gaia} accurately covers a full range in parameters such as absolute extinction, differential extinction, distance, metallicity, and age. In particular, due to the different cuts on \emph{Gaia} data used in this work, we often detect significantly more member stars for many clusters and up to two magnitudes fainter than the membership lists of \cite{cantat-gaudin_clusters_2020}; hence, particularly for more distant OCs, our membership differences have a significant impact on inferred parameters, making existing literature catalogues inappropriate to use as training data. Simulated data, if it can be simulated accurately enough, would offer an attractive way to quickly generate new training data applicable to new methodologies and new \emph{Gaia} datasets or even other instruments, entirely based on a ground truth or `best estimate' of how OCs should appear based on prior knowledge from stellar evolution models. Additionally, training data based on real clusters are biased towards an unknown selection effect of how a human defines a real cluster -- whereas for simulated data, we are able to exactly state the distributions we assume real OCs are drawn from, hence giving more knowledge of any selection biases this may cause.

A key issue found in early experiments is that typical machine learning approaches are deterministic, and hence do not quantify the underlying uncertainties on their predictions. To aid with the use of simulated data, we adopt an approximate Bayesian neural network (BNN) framework using variational inference. In practice, true Bayesian machine learning is impractical to achieve with current methods; however, variational inference-based approaches offer an approximate and fast way to estimate the uncertainty of a neural network model by approximating parameters with simple probability distributions \citep{goan_bayesian_2020, jospin_hands-bayesian_2022}, of which networks can then be sampled multiple times to produce a probability distribution for their output. The BNN approach we trialed had similar accuracy to a purely deterministic one except while also outputting uncertainties, allowing us to estimate the uncertainty of our classifier. We provide a broader overview of our adopted variational inference-based approach in Appendix~\ref{app:bayesian_nets}. Next, we discuss the creation of training data for our CMD classifier.



\subsection{Simulated real OCs}\label{sec:cmd_classifier:tps}

A number of steps were used to generate examples of real OCs to train our CMD classifier. Basic OC generation was conducted using SPISEA \citep{hosek_jr_pypopstar_2020} to simulate single-population clusters from PARSEC evolution models \citep{marigo_new_2017}, with extinction calculated star-by-star using a \cite{cardelli_relationship_1989} extinction law with $R_V=3.1$. Stars were sampled from these isochrones with SPISEA using a \cite{kroupa_variation_2001} IMF. In addition, SPISEA was used to supplement simulated OC CMDs with unresolved binary stars based on general relations derived in \cite{lu_stellar_2013} for zero-age star clusters. The values in this work were found to correspond relatively well to \emph{Gaia} observations, with a mass-dependent multiplicity frequency peaking at 100\% for clusters of masses above $5~M_{\sun}$. In practice, unresolved binary stars have negligible impact on the final cluster CMDs fed to the network, as typical binary sequences observed in \emph{Gaia} photometry are smaller than the size of the pixels in input CMD images. SPISEA was also used to apply Gaussian-distributed differential reddening, with values up to a standard deviation of 0.6 in the highest cases, reflecting the most extreme examples of differentially reddened reliable clusters found in \cite{cantat-gaudin_clusters_2020}.

Next, a random location on the galactic disk was selected for each cluster, which was used to simulate a realistic selection function and photometric errors. The magnitude-dependent selection function of \emph{Gaia} DR3 at each given location was queried using the \texttt{selectionfunctions} package presented in \cite{boubert_completeness_2020} and \cite{boubert_completeness_2020-1}, which gives the basic probability that a source appears in \emph{Gaia} as a function of position and G-band magnitude. We use the online version of their package updated for \emph{Gaia} DR3. The \texttt{selectionfunctions} package is based on the \texttt{dustmaps} package from \cite{green_dustmaps_2018}. In addition, the selection function of every cluster was also corrected for the cuts to \emph{Gaia} data applied in Sect.~\ref{sec:data:outliers}. During the preparation of this work, \cite{cantat-gaudin_empirical_model_2023} released a new selection function for \emph{Gaia} DR3 which suggested that the earlier work of \cite{boubert_completeness_2020, boubert_completeness_2020-1} can be over-confident at the faint end; however, given that our cluster membership lists are overwhelmingly dominated by the selection function of our cuts on \emph{Gaia} data at magnitudes $G>18$, and not the pure selection function of \emph{Gaia}, we found that it made too small of a difference to our simulated clusters to be worth updating our training data for, although we will adopt their work in future works. Realistic photometric uncertainties were added to sources based on the distribution of source uncertainties at the selected location, which are generally larger in crowded fields. We added systematic offsets in simulated BP and RP \emph{Gaia} photometry for faint sources using relations in \cite{riello_gaia_2021}. 

Outliers were not added to simulated cluster CMDs, as most clusters are already detected with very few or no outliers; instead, we wish the CMD classifier to quantify the evidence for a cluster being real based on its photometry alone, which photometric outliers inherently reduce. In this way, CMDs of clusters with a high number of outliers are scored more negatively by the network as they have less photometric evidence supporting them being real. Blue stragglers were also not added to cluster CMDs as they are indistinguishable from photometric outliers, although in practice, real OCs with blue straggler stars were not found to be scored significantly lower by the trained network.


10\,000 examples of simulated real clusters were generated to use as one half of the simulated cluster dataset. Distributions of parameters such as age $\log t$, extinction $A_V$, differential extinction $\Delta A_V$ and distance modulus $m-M$ were carefully chosen after many iterations to minimise systematics deriving from the overall distribution of training data in the dataset, while ensuring that the CMD classifier was trained on a representative set of simulated real OCs. Fundamentally, the objective of the training data are not to match the real distribution of OCs, but rather to yield an unbiased and representative sample of OCs to train the BNN on, such that the BNN can provide an unbiased classification of any object. For instance, while a distribution of the number of visible stars $n$ based on the distribution of stars in \cite{cantat-gaudin_clusters_2020} (corrected for our deeper magnitude limit) was found to work well to produce an unbiased classifier, in other cases, such as for $\log t$ and $m-M$, the use of a uniform distribution (instead of one based on the expected distribution of clusters) was essential to avoid biasing the classifier towards certain ages or distances. These distributions are listed in Table~\ref{tab:simulated_oc_distributions}.

\subsection{Simulated fake OCs}\label{sec:cmd_classifier:fps}

A number of methods to simulate fake OCs reminiscent of false positives sometimes reported by HDBSCAN were trialed. As a clustering algorithm, the member stars of each cluster reported by the algorithm are spatially correlated, with a similar position, proper motion, and parallax. Hence, it is important that false positives contain member stars with similar astrometric parameters. Simply randomly selecting stars from \emph{Gaia} data to construct each false positive was found to result in clusters that were too pessimistic.

Instead, to generate false positives with spatially correlated member stars, a star was first selected randomly from the entire \emph{Gaia} dataset as an origin point. This ensures inherently that false positives are more likely to occur in the densest regions of the \emph{Gaia} dataset, which was a behaviour observed inherently for HDBSCAN in Paper~1. A total number of stars for the cluster was selected from the same distribution as used for simulated real OCs. Then, a 5D hypersphere in position, proper motion, and parallax was expanded randomly around this star until the hypersphere contained the required number of stars. In this way, false positives with spatially correlated member stars were generated. Actual OCs make up a small enough portion of the \emph{Gaia} dataset -- $610 \, 000$ in the final version of the catalogue, or fewer than 0.1\% -- that it was not found to be necessary to first remove them from data used to generate false positives. This is similar to the false positive generation method used in \cite{castro-ginard_hunting_2022}.

10\,000 false positives were generated using this methodology to provide the other half of the training dataset. While most false positives have obviously poor quality CMDs, false positives generated from regions of field stars with roughly homogeneous ages and composition (such as from the galactic halo) often had more homogeneous CMDs, that could be compatible with highly differentially reddened OCs. However, this is a useful property of the training dataset, given the variational inference approach used in the network: this `overlap' between highly differentially reddened true positives and chance alignments of somewhat-similar field stars reflects on the real distributions of field stars in the galactic disk. Real \emph{Gaia} cluster candidates with worse-quality CMDs making them compatible with both a real OC or a chance clustering of field stars hence have broad or bi-modal PDFs from the BNN CMD classifier, reflecting how photometry alone offers only poor evidence of whether or not these objects are real or fake star clusters.


\subsection{Test dataset}\label{sec:cmd_classifier:test_data}

\begin{table}
\caption{Human classifier performance.}
\centering
\label{tab:human_classifier}
\begin{tabular}{l c | c c c c}
\hline\hline
 & & \multicolumn{4}{c}{Percent classified as} \\
Dataset & Size & TP & TP? & FP? & FP \\
\hline
Test data          & 2000 & 53.6 & 26.5 & 11.0 & 8.9 \\
Simulated real OCs & 250  & 72.0 & 20.0 & 6.0  & 2.0 \\
Simulated fake OCs & 250  & 14.0 & 26.8 & 28.0 & 31.2 \\
\hline
\end{tabular}
\tablefoot{Results of human classification when applied to a test dataset of 2000 clusters detected by HDBSCAN in this work as well as two datasets of simulated real and fake clusters.}
\end{table}

In order to test the trained networks against real \emph{Gaia} data and ensure that they can be generalised from their training on simulated data to use on real data, a test dataset of 2000 clusters randomly selected from the initial HDBSCAN clustering was selected and classified by hand, in addition to 250 simulated real clusters and 250 simulated fake ones to estimate the accuracy of human classification. These different datasets were classified in one classification run to avoid biasing the human classifier. Clusters were classified into `true positive' (TP) and `false positive' (FP) categories, in addition to two other categories for clusters that are most likely to be true or false clusters but are somewhat uncertain (abbreviated as `TP?' or `FP?'), due to the presence of outliers, a small number of stars, or very high differential reddening that is compatible with both an association of field stars or a highly differentially reddened OC. The results of this classification are shown in Table~\ref{tab:human_classifier}.

Of clusters reported by HDBSCAN, 53.6\% were hand-classified as being highly likely to be real, with a further 26.5\% being potentially real, suggesting that most clusters we detect have a reliable CMD. Only 8.9\% were highly unlikely to be real with a further 11.0\% classified as probably not real, suggesting that around 80\% of clusters reported by HDBSCAN are likely to have single stellar populations based on human classifications. 

In testing the human classifier, 92.0\% of simulated real clusters were correctly classified as real or potentially real, although only 59.2\% of simulated fake clusters were classified as false or potentially false. 14.0\% of simulated fake clusters were in fact classified as highly likely to be real. This shows the inherent limitations of using photometry to validate OCs, as spatially correlated groups of field stars can often have somewhat-homogeneous CMDs when all field stars in a given region have a similar age and chemistry (see Sect.~\ref{sec:cmd_classifier:fps}), which can even fool a human classifier. This is particularly common in the halo and thick disk where most stars have a similar, old age. This is an important limitation of the human-classified test data to bear in mind, as a small fraction of clusters classified by hand as true positives will always in fact be false positives. Nevertheless, CMD classification is still a necessary validation tool to help ensure that detected cluster candidates are reliable, as many of the worst quality clusters can still be removed with this method.


\subsection{Network training and validation}\label{sec:cmd_classifier:train_and_val}

\begin{figure}[t]
   \centering
   \includegraphics[width=\columnwidth]{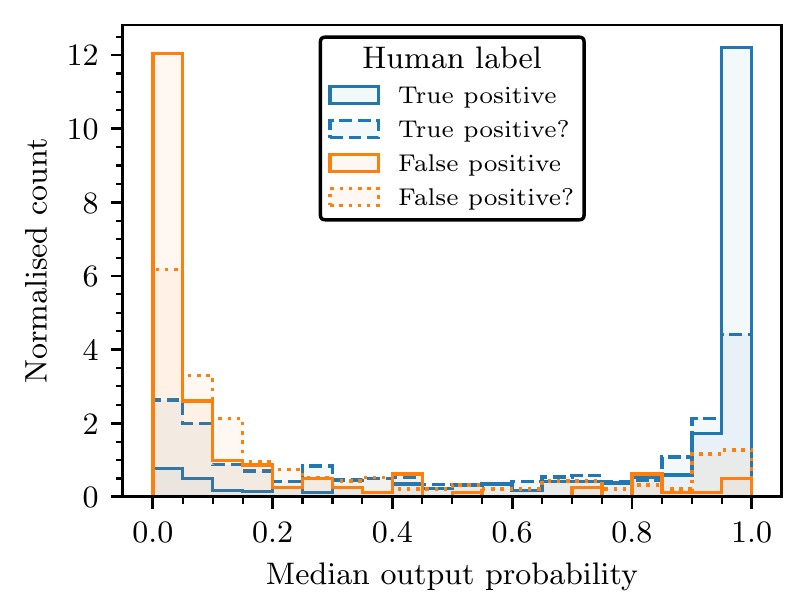}
   \caption{Performance of the CMD classifier on the independent test dataset of 2000 clusters detected by HDBSCAN in \emph{Gaia} data and labelled by hand. Clusters are labelled as true positives or false positives, with clusters where the human classifier was less certain being additionally flagged.}%
   \label{fig:classifier_test_results}
\end{figure}

\begin{figure*}[t]
   \centering
   \includegraphics[width=\textwidth]{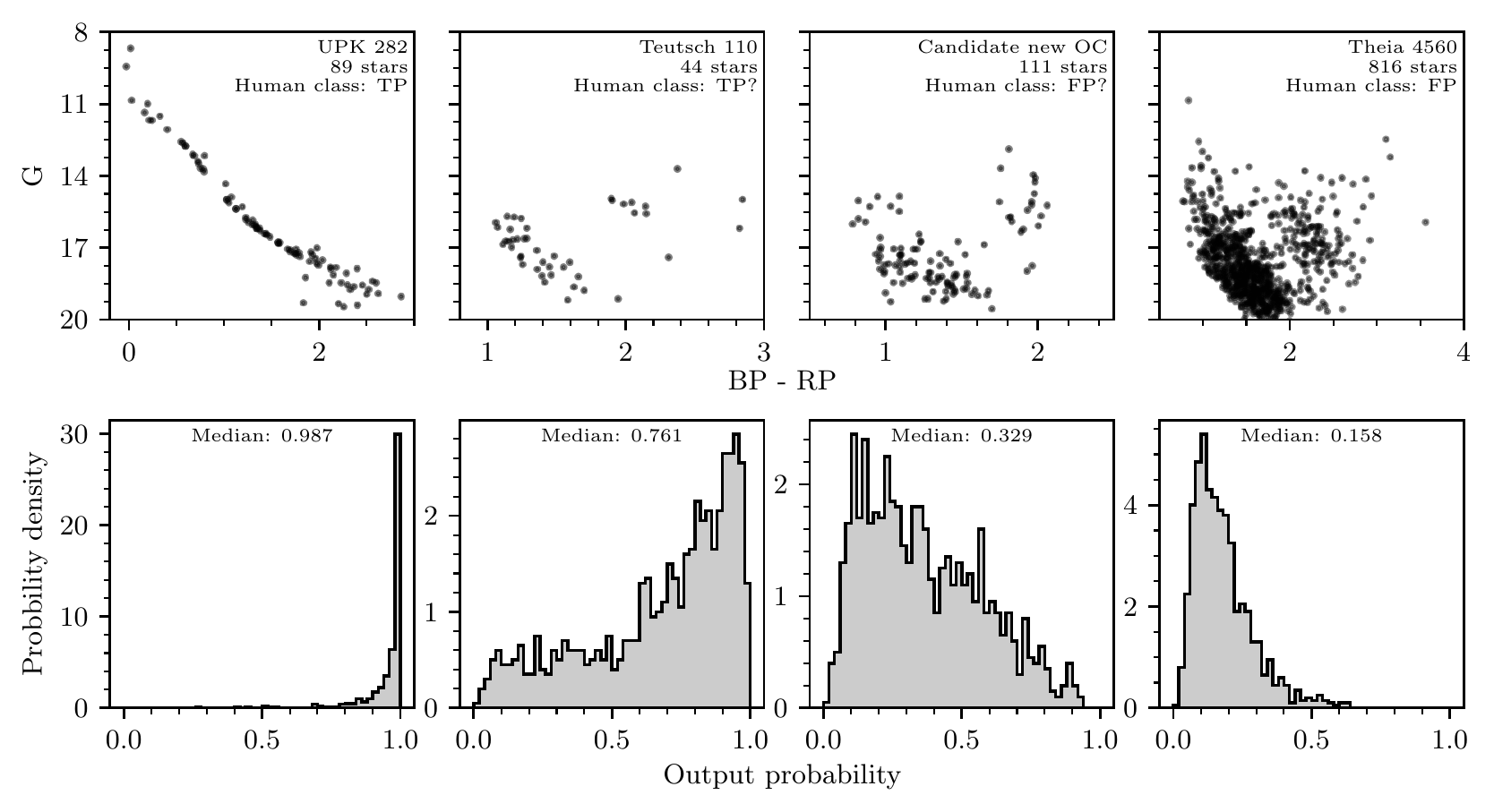}
   \caption{Four examples of classified cluster CMDs from the test dataset, with cluster CMDs on the top row and their PDFs of predicted probabilities on the bottom row. Cluster names and human-assigned labels are indicated on the figures. PDFs are generated by sampling the CMD classifier 1000 times for every cluster.}%
   \label{fig:classifier_examples}
\end{figure*}

The $20\,000$ simulated real and fake OCs were split randomly into a training set of $16\,000$ clusters and a validation dataset of $4\,000$ clusters to assess network overfitting. As the simulated fake OCs have a different distribution of distance moduli to the simulated real OCs, fake OCs at undersampled and oversampled distances were weighted to be emphasised more or less strongly during training, preventing systematics due to differences in distance distributions.

We used the implementations of neural networks and probabilistic layers in TensorFlow \citep{abadi_tensorflow_2015, abadi_tensorflow_2016} and TensorFlow Probability \citep{dillon_tensorflow_2017} for all networks used in this work. Networks were trained with the Adam optimisation algorithm \citep{kingma_adam_2017}. A number of different neural network structures were trialed. Convolutional neural networks (CNNs), which convolve two-dimensional input with learnt filters, were found to perform ideally for the problem at hand, and have seen extensive use in the astronomical literature \cite[e.g.][]{castro-ginard_hunting_2022, becker_cnn_2021, killestein_transient-optimised_2021}.

As input, the optimal network trialed used cluster CMDs converted to absolute magnitudes, with stars of absolute G magnitudes greater than 10 or lower than $-2$ cut away. Generally, this cuts certain very low mass M stars and bright O stars from cluster CMDs, which were found to be poorly simulated by PARSEC isochrones with their inclusion only worsening network performance on real data. In practice, very few stars are cut due to this limitation, with O stars making up only a very small proportion of sources in young clusters and M dwarfs fainter than $M_G=10$ only being brighter than $G=20$ for clusters within 1~kpc, at which point the rest of the cluster CMD can be resolved well. In addition, $BP-RP$ colours were cut between -0.4 to 4, which in practice is a wide enough colour range to include almost all sources but while providing a good range to discretise cluster CMDs between. Sources with very low BP and RP fluxes that have overestimated BP or RP magnitudes were removed using cuts from \cite{riello_gaia_2021}, as these also only confused the network, despite these systematics being simulated in the training data. Finally, in terms of structure, the optimal network trialed was trained on CMDs discretised into $32\times32$ pixel images, corresponding to pixels of size $0.38\times0.11$~mag. These images were first processed by three convolutional layers with $5\times5$ pixel kernels of 6, 16, and 120 filters respectively. Max pooling layers were placed between these convolutional layers to speed up training and inference. Convolution layer output was connected to a single densely connected layer of 128 nodes, with a final single node for output. The distance modulus of the cluster based on the parallax-derived cluster distances was also fed to the network as an auxiliary input into the 128 node dense layer, in a similar way to the network of \cite{cantat-gaudin_painting_2020} which also uses both photometric and astrometric input simultaneously. All layers used Rectified Linear Unit (ReLU) activation other than a sigmoid activation function applied to the final output to constrain network output in the range $[0,1]$ as a probability distribution.

The final network had binary accuracies (the percentage of clusters given the correct true or false label) of 95\% for both training and validation data, indicating that the network did not overfit to training samples when compared with other simulated data. Fig.~\ref{fig:classifier_test_results} shows the performance of the network compared to the human-labelled test dataset of real clusters detected by HDBSCAN in \emph{Gaia} after sampling the network 1000 times to generate PDFs for every object, with 85.5\% of clusters labelled highly likely to be real and 91.3\% of clusters labelled highly unlikely to be real having a median predicted probability greater or less than 0.5 respectively. Clusters where the human classifier was less certain have a much broader distribution, although this also reflects inherent uncertainties in the test dataset discussed in Sect.~\ref{sec:cmd_classifier:test_data}. Finally, only 4.3\% and 2.5\% of highly likely real and highly likely false clusters had predicted labels that disagree with human labels at more than the 2$\sigma$ level -- namely, that 97.5\% of their PDF is below or above 0.5 respectively. It is important to recall that these quantities merely validate the general agreement between two independent classifiers (the human classifier and the automated CMD classifier) on the same dataset, and do not exactly measure the ground truth sensitivity or accuracy of the CMD classifier, as the human class labels themself are uncertain Sect.~\ref{sec:cmd_classifier:test_data}. Instead, these data show that the CMD classifier can perform comparably well to human classification, except with the added bonuses of speed and reproducibility.

Fig.~\ref{fig:classifier_examples} shows CMD classifier PDFs for four clusters from all human classes, including the names of any clusters that crossmatched to real objects. In general, CMD classifier predictions generally agreed well with the human-assigned labels, also generally with higher uncertainty and a broader PDF in cases where the human classifier was less certain. For clusters with clear, high-quality CMDs such as UPK~282, the CMD classifier outputs PDFs that strongly suggest they are real. Teutsch~110 is a less well-defined cluster that, if real, must have differential reddening and a few outliers, and is hence not classified as strongly. The candidate new cluster shown is a similar case albeit with a worse CMD, making it relatively unlikely to be real given this HDBSCAN detection. Finally, Theia~4560 is visible as a large and statistically significant overdensity in \emph{Gaia} data as detected by \cite{kounkel_untangling_2020}, although the overdensity as detected in this work does not appear to contain a homogeneous population of stars and is hence classified weakly. CMD classifier median probabilities and confidence intervals for all clusters are listed in Table~\ref{tab:catalogue}, based on 1000 samples of the network for each cluster.


\section{Age, extinction, and distance inference}\label{sec:agenn}
\subsection{CMD classifier modifications}

\begin{figure*}[t]
   \centering
   \includegraphics[width=\textwidth]{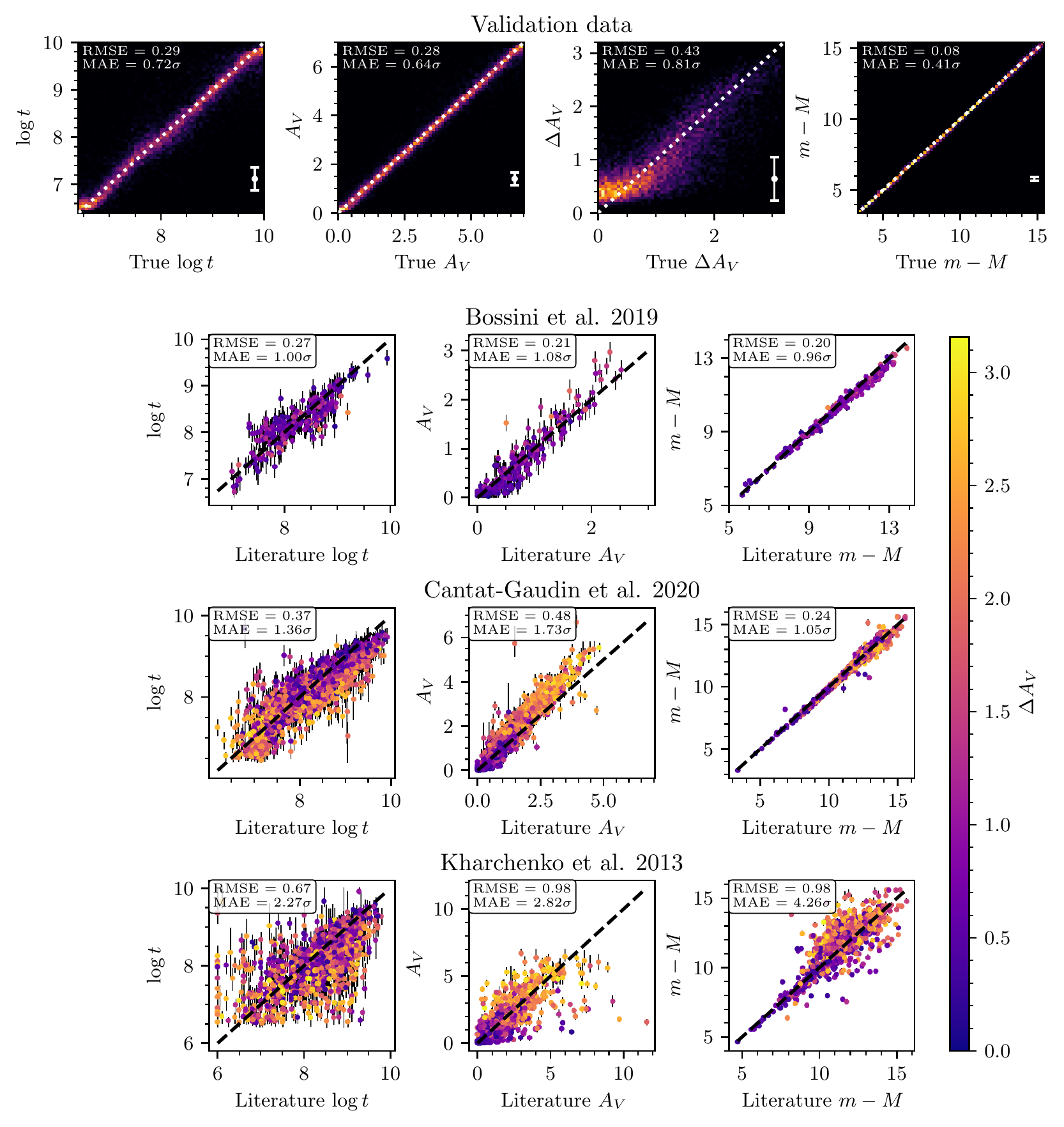}
   \caption{Photometric parameters derived in this work compared against test datasets. \emph{Top row}: 2D histograms showing the performance of the trained photometric parameter inference network on all $10\,000$ clusters from the validation dataset. The mean output uncertainty is shown with white error bars. As indicated by the dashed lines, predicted values on the $y$ axis should be equal to true values on the $x$ axis. The root mean square error (RMSE) and mean absolute error in terms of output network uncertainty (MAE) are given in the top left. All plots and the RMSE are in units of magnitude other than on age plots which are logarithms of cluster age in years. \emph{Other rows}: comparison between network predicted parameters and ages, extinctions, and distance moduli for 247, 1753, and 1206 clusters in common with the catalogues of \cite{bossini_age_2019}, \cite{cantat-gaudin_painting_2020}, and \cite{kharchenko_global_2013} respectively. Points are shaded based on the differential extinction we infer for each cluster.}
   \label{fig:agenn_comparison}
\end{figure*}

\begin{figure}[t]
   \centering
   \includegraphics[width=\columnwidth]{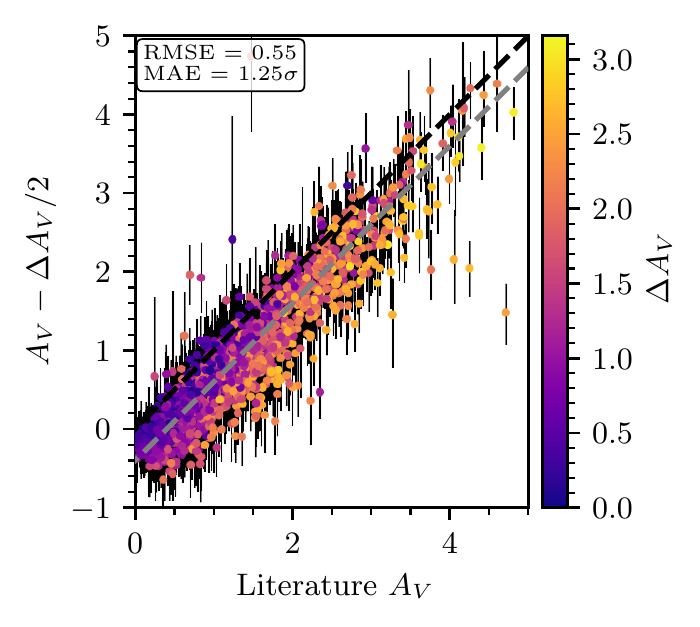}
   \caption{Extinction values from \cite{cantat-gaudin_painting_2020} compared against this work when corrected for differential extinction with an estimate of cluster differential extinction, plotted in the same style as Fig.~\ref{fig:agenn_comparison}. The dashed black line shows where $y$ values equal $x$ ones; the dashed grey line shows the same but offset by -0.4.}%
   \label{fig:agenn_cantat_gaudin_comparison}
\end{figure}

\begin{figure*}[t]
   \centering
   \includegraphics[width=\textwidth]{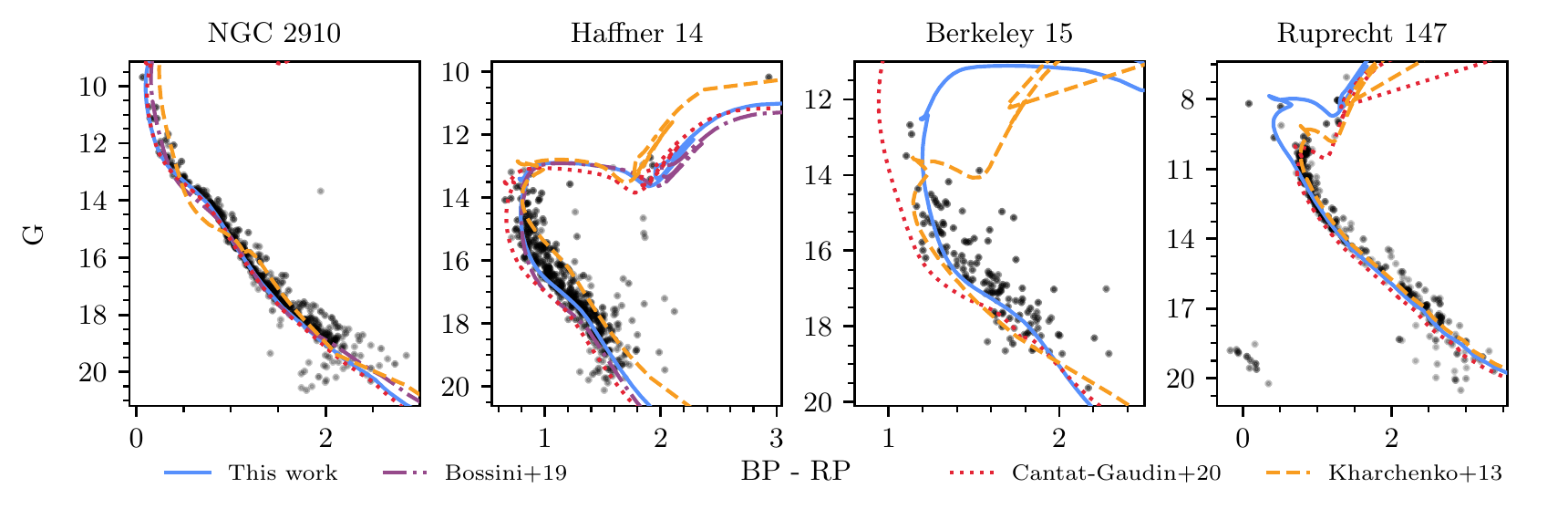}
   \caption{Predicted cluster isochrones from this work (solid blue line) compared with those from other works. Cluster members are plotted in black and shaded according to their membership probability.}%
   \label{fig:agenn_fits}
\end{figure*}

While not a main focus of this work, we also show that the approach based on simulated data and an approximate BNN using variational inference is also applicable for age $\log t$, extinction $A_V$, differential extinction $\Delta A_V$ and distance modulus $m-M$ inference. Recently, \cite{cantat-gaudin_painting_2020} use a neural network to infer $\log t$, $A_V$ and $m-M$ for around 2000 OCs. In their work, a training dataset based on simulated OCs alone is not found to be sufficiently accurate to train a neural network. While simulated data were found to be accurate enough for the CMD classifier in Sect.~\ref{sec:cmd_classifier}, parameter inference is more challenging, as a network must learn to infer multiple parameters from a CMD alone and generalise this accurately to real data. However, our approach has a number of differences to theirs: firstly, we use a convolutional neural network, which may be better able to capture structure in CMDs due to its 2D approach, which may also reduce training data overfitting; secondly, our network is approximately Bayesian, and includes uncertainty estimates that quantify when it may have failed; finally, although \cite{cantat-gaudin_painting_2020} do not elaborate on how they simulate clusters in their work, our methodology is be different and may produce different results. Hence, despite recent literature suggesting that using purely simulated data is not possible for parameter inference with CMDs, it is still worth attempting, as training on simulated data is attractive for reasons discussed in Sect.~\ref{sec:cmd_classifier}.

To create a parameter inference network, we used a similar network structure to that of Sect.~\ref{sec:cmd_classifier:train_and_val}, except with some tweaks to the network output to infer parameters. To better predict the aleatoric uncertainty of network output for this multiple-parameter network, network output was changed to a beta distribution for each parameter. These distributions can take any shape from a uniform (completely uncertain) distribution to a single point-like estimate. The output was then scaled to be within the minimum and maximum ranges of the training data. To train the network, 50\,000 simulated clusters were created using the same methodology as in Sect.~\ref{sec:cmd_classifier:tps}, changing the distribution of cluster extinctions $A_V$ (as defined in Table~\ref{tab:simulated_oc_distributions}) to simply be uniform between 0 and 7.

In initial comparisons with literature results, differential reddening was found to strongly correlate with disagreements in extinction (and to a lesser extent, age) between this work and others. A primary cause of this is that while many works \citep[e.g.][]{cantat-gaudin_painting_2020, bossini_age_2019} use the so-called `blue edge' of a CMD for isochrone fitting, meaning that $\Delta A_V$ is only positive. This contrasts to SPISEA's default $\Delta A_V$ model, which is Gaussian -- with cluster stars having both positive and negative $\Delta A_V$ values. 

However, changing SPISEA's $\Delta A_V$ model to also only be positive (and hence defining $\Delta A_V$ in terms of the blue edge of cluster CMDs) was not found to be helpful. Owing to HDBSCAN's high sensitivity, we detect a higher number of stars outside of the core of clusters than in the membership lists of \cite{cantat-gaudin_clusters_2020}, which are constructed with the UPMASK algorithm \citep{krone-martins_upmask:_2014} and for many clusters only select stars in the core. This means that our CMDs are constructed from clusters with significantly larger angular extents on the sky and are hence often more strongly differentially reddened than in \cite{cantat-gaudin_clusters_2020}, with many clusters having a blue edge at an extinction value up to 1 magnitude lower than in \cite{cantat-gaudin_clusters_2020}. For instance, NGC~884 is an example of this, with our membership list being larger and more strongly differentially reddened. A blue-edge based definition of $A_V$ means that different works produce different values of $A_V$ depending on how sensitive their membership recovery process is.

Instead, we continue using the default SPISEA $\Delta A_V$ definition centred on the mean cluster $A_V$, but while also using the network to infer $\Delta A_V$ for every cluster, which can then be used as a correction to convert between extinctions in this work and others that use a blue-edge definition. In practice, $\Delta A_V$ is very difficult to measure, as it is degenerate with other effects that broaden cluster CMDs, including unresolved binary stars and outliers. Against validation and test data, our median $\Delta A_V$ values are found to be offset by around 0.4 due to unresolved binaries. Nevertheless, this parameter is helpful to aid comparisons with literature works.

Finally, we also updated our $\Delta A_V$ model from the Gaussian default model in SPISEA to instead use the differential reddening as would be expected from stars sampled from a King profile \citep{king_structure_1962}, assuming a first order (linear) gradient in differential extinction across a cluster. This model is narrower than the Gaussian model while retaining highly differentially reddened stars (which would be at the outskirts of a cluster), and was found to slightly improve $\Delta A_V$ inference. This model depends on two parameters: the total differential extinction across a cluster, which was matched to have the same range as the previous Gaussian model at a $3\sigma$ level; and the ratio between core and tidal radius, which was set to the median value for open clusters from \cite{kharchenko_global_2013}.

Against our validation dataset of $10\,000$ simulated clusters, the network performs well with no clear systematics in $\log t$, $A_V$ or $m-M$. However, owing to the degeneracy between $\Delta A_V$ and other effects such as unresolved binary stars, outliers, and photometric uncertainties, values of $\Delta A_V$ smaller than 0.4 are not typically correctly predicted, although the true value is typically still within 1$\sigma$ uncertainty of the predicted value. These results are plotted on the top row of Fig.~\ref{fig:agenn_comparison}.

 Using the best trained network after a number of experiments, all clusters in our catalogue closer than a maximum distance of 15~kpc have ages, extinctions, differential extinctions, and distance moduli listed in Table~\ref{tab:catalogue}. These parameters are based on 1000 samples of the network for each cluster.

\subsection{Comparison with other works}

 We briefly compare our photometric parameters to other works in the literature. Firstly, Fig.~\ref{fig:agenn_fits} shows example predicted isochrones for four OCs in this work. In the first case, NGC~2910 is a cluster with a well-behaved isochrone where all works agree relatively well. On the other hand, Haffner~14 shows relatively strong differential reddening, and different definitions of differential reddening between different works cause isochrone fits to disagree. Berkeley~15 is a sparse cluster where both differential reddening and field star outliers affect different works in different ways, with our updated \emph{Gaia} DR3 membership list having fewer outliers than that of \cite{cantat-gaudin_characterising_2018}. Ruprecht~147 is a nearby and particularly old cluster ($\sim1$~Gyr), where blue straggler stars systematically affected our network and caused an incorrect younger age value to be predicted for this cluster. It is clear from these plots that for all but the most well-behaved OCs, different works can have different photometric parameters.

Fig.~\ref{fig:agenn_comparison} compares all network predictions with values from four test datasets. An advantage of our simulated training approach is that network predictions can now be compared to other literature works, which act as independent test datasets which can verify the accuracy of our network. It is important to note that our results never agree perfectly, however, particularly since all works we compare to are based on \emph{Gaia} DR2 or pre-\emph{Gaia} OC membership lists that may be significantly less clean or have significantly fewer stars than our \emph{Gaia} DR3 membership lists.

\cite{bossini_age_2019} provide a catalogue of precise OC parameters from Bayesian isochrone fitting using the BASE-9 algorithm \citep{hippel_inverting_2006}. A key difference is that their work uses metallicity estimates from the literature where available, whereas our approach is based entirely on \emph{Gaia} DR3 parameters and assumes a given cluster can have any metallicity as drawn from a broad probability distribution based on literature values (Table~\ref{tab:simulated_oc_distributions}). Nevertheless, our results still agree well with theirs in $\log t$, $A_V$ and $m-M$. In cases where our $\log t$ estimates disagree most strongly, this is typically due to differences in OC membership list. There is however a possible minor systematic between our two works for OCs with extinctions below 0.6, many of which we infer smaller extinctions for than them; this may be as a result of $A_V$ vs. metallicity degeneracies. However, their values are typically only 1 to 2$\sigma$ from ours.

Our parameters agree less strongly with the results of \cite{cantat-gaudin_painting_2020}, which are derived from a neural network trained on isochrone fits from a variety of works \citep[including][]{bossini_age_2019}. This is to be expected to some extent, as while \cite{bossini_age_2019} only fit isochrones to a subset of OCs with clean membership lists and the least differential reddening, \cite{cantat-gaudin_painting_2020} fit isochrones to all known OCs at the time, including many sparse objects which may now have significantly different membership lists in our current \emph{Gaia} DR3 work. However, some differences persist. A clear systematic in our and their $A_V$ values is clear, although this is likely due to their different blue edge definition of extinction (whereas our network fits to the mean extinction in a cluster.) Figure~\ref{fig:agenn_cantat_gaudin_comparison} shows a crude conversion between our $A_V$ values and their blue-edge $A_V$ values. While this removes the systematic difference in gradient, our converted $A_V$ values are still generally smaller than theirs by around 0.4 to 0.5 on average. This is likely due to two effects; firstly, as shown by the results on validation data, $\Delta A_V$ is generally overestimated for our validation data by around $\sim0.4$ due to degeneracies with unresolved binary stars, outlier non-member stars, and photometric uncertainties, which may explain some of this discrepancy, particularly for clusters with lower $\Delta A_V$ values. Secondly, our membership lists generally cover a wider extent on the sky than those used in \cite{cantat-gaudin_painting_2020}, meaning that our clusters are often larger and hence are more extremely differentially reddened between separate sides of the cluster; hence, a conversion between the works based on our $\Delta A_V$ values is likely to frequently over-correct for the difference in $A_V$ definition. Finally, some of our ages for the oldest clusters ($\log t > 9$) appear systematically younger, on average by around $2\sigma$; in some cases, this may be due to our fits being disrupted by blue straggler stars (Fig.~\ref{fig:agenn_fits}, see Ruprecht~147.) The training data we use for our photometric parameter inference are adapted from our CMD classifier in Sect.~\ref{sec:cmd_classifier}, for which blue straggler stars were not found to have a negative impact on the accuracy of our network and were hence not included. Future works using purely simulated data to train a photometric parameter inference neural network would benefit from inclusion of blue straggler stars in their training data, although in practice the origin of blue stragglers is still disputed, and these stars may hence be challenging to simulate accurate photometry for \citep{boffin_ecology_2015, cantat-gaudin_milky_2022}.

Finally, our results have limited agreement with those of \cite{kharchenko_global_2013}. While some clusters have similar values between their work and ours, particularly for $A_V$ and particularly for the largest and most clearly defined clusters (Fig.~\ref{fig:agenn_fits}), many sparse clusters that were difficult to detect before \emph{Gaia} have very different photometric parameters. This typically appears to be caused by extremely different  cluster membership lists. Before \emph{Gaia}, OCs were often challenging to separate from field stars \citep{cantat-gaudin_milky_2022}, requiring that suspected outliers be removed iteratively to improve CMD quality \citep{kharchenko_global_2012}. However, this process can also remove true cluster members, which can cause resulting cluster membership lists to be incorrect \citep{cantat-gaudin_clusters_2020}. This discrepancy with the results of \cite{kharchenko_global_2013} is also reported by \cite{cantat-gaudin_painting_2020}, who also find that many photometric parameters derived before \emph{Gaia} are strongly discrepant with current results. In addition, while the number of member stars reported in \cite{kharchenko_global_2013} is generally a poor predictor for whether or not a given cluster in their work has very different parameters to ours, there are some cases (such as clusters in their work with $A_V>5$ that we derive much smaller values for) where the most discrepant clusters were also the smallest, with fewer than 20 member stars in reported in \cite{kharchenko_global_2013}.

Although approximate, these results still agree well within the sample-limited but accurate Bayesian isochrone fits of \cite{bossini_age_2019} and agree relatively well (albeit with some caveats) with the machine learning derived parameters of \cite{cantat-gaudin_clusters_2020}. This work offers a large and homogeneously derived catalogue of photometric parameters with sufficient accuracy for basic analysis. In the next section, we use the ages and extinctions we derived here to aid with discussion of our cluster sample.








\section{Crossmatch to existing catalogues}\label{sec:crossmatching}

\subsection{Crossmatch strategy}

Before conducting further analysis on the cluster catalogue, such as restricting it to only clusters with reliable colour-magnitude diagrams or removing moving groups, it is helpful to crossmatch our results to literature catalogues to allow for easier comparisons between derived parameters and other works. In particular, this makes it possible to compare whether clusters reported in other works are compatible with real open clusters given further parameters derived in Sect.~\ref{sec:cmd_classifier} and the third paper in this series, Hunt~\&~Reffert,~\emph{in~prep.}, where we will derive dynamical parameters for our census of star clusters.

In Paper 1, we crossmatched by assigning matches to clusters when their mean positions were compatible to within their tidal radii and when their mean proper motions and parallaxes were compatible within five standard errors. In initial testing, the crossmatch strategy of Paper 1 was found to be insufficient for two reasons when comparing between \emph{Gaia} DR3 astrometry and \emph{Gaia} DR2 astrometry, in addition to a further issue with the positional strategy used.

Firstly, the standard errors on mean proper motions and parallaxes in \emph{Gaia} DR2 can be as small as 5 to 10~$\mu$as for the largest clusters in catalogues such as \cite{cantat-gaudin_clusters_2020}, although this is smaller than estimated upper limits on systematics in \emph{Gaia} DR2 of 50~$\mu$as \citep{lindegren_gaia_2018}. Many reliable clusters are hence missed when treating DR2 positions exactly, as they have systematics significantly larger than their standard errors, with positions in DR3 that can deviate systematically from their DR2 positions by 50~$\mu$as or more.

Secondly, membership lists can differ between works and can be significantly different for the same cluster -- for instance, works such as \cite{castro-ginard_hunting_2020} only used stars down to $G=17$, whereas this work often has membership lists down to $G\sim20$. Many clusters hence have significantly different membership lists that can result in different mean parameters, particularly for asymmetric clusters.

Our positional crossmatch strategy was also revised and improved. Paper 1 used a conservative strategy for matching on position, which assumed that a cluster is a positional match if the centre of the literature cluster is closer than either the Paper 1 or literature radius for a given cluster. However, in practice, this strategy appears almost always too conservative, as many distant, compact clusters reported in catalogues such as \cite{froebrich_systematic_2007} would match to large, nearby clusters that happen to contain the distant object within one radius, despite the cluster centres being strongly incompatible given the smaller (literature) radius.

To improve positional crossmatching, we instead define a positional match to require that the centre of the literature cluster is closer than both the current and literature radius, which in almost all cases still recovers reliable matches but while not erroneously matching to compact, distant objects with significantly different sizes and cluster centres. Then, for catalogues with \emph{Gaia} astrometry available, we also match on proper motions and parallaxes, requiring that the new mean proper motion and parallax are within two standard deviations of the literature value (with both current and literature standard deviations summed in quadrature.) This approach with standard deviations matches clusters if a new cluster is within allowed ranges of the dispersion of the current and literature entries, with the principles that exact statistical matching based on standard errors is not possible as unknown systematic errors dominate, and that a cluster within the dispersion of a literature entry is likely to be the same object. Using a higher maximum value of the dispersion was not found to significantly increase the number of literature clusters recovered by more than 1\%, but while adding many false crossmatches to other nearby objects that greatly worsen the reliability of the overall crossmatching process.

Some special cases are also worth mentioning: the catalogue of \cite{kharchenko_global_2013} is based on PPMXL proper motions and distances from isochrone fitting by hand, which are generally significantly less accurate than \emph{Gaia} astrometry. Hence, we crossmatch to \cite{kharchenko_global_2013} with both a position-only and a second positions, proper motions, and distances crossmatch which can more strongly confirm the most reliable matches. Some catalogues list only a radius containing 50\% of members for entries \citep[e.g.][]{cantat-gaudin_clusters_2020}; for these catalogues, we use twice this radius to approximate the total size of the cluster. Other works \citep[e.g.][]{castro-ginard_hunting_2020, he_new_2022} list only standard deviations of the mean position; for these catalogues, we use twice the geometric mean of this standard deviation on position to approximate the total size of the cluster. Finally, \cite{kounkel_untangling_2020} does not list uncertainties or dispersions on mean parameters, and so these were manually recalculated with our own pipeline using their lists of members.

After an extensive search of the literature for recent catalogues, excluding works already listed entirely in other catalogues (such as \cite{froebrich_systematic_2007}, which appears in its complete form within \cite{bica_multi-band_2018}), we crossmatch against 26 different works listed in Table~\ref{tab:crossmatches}. In addition, as our catalogue contains many moving groups, globular clusters, and a handful of clusters associated with the Magellanic clouds, we also crossmatch against the \cite{kounkel_untangling_2020} catalogue of predominantly moving groups, the \cite{vasiliev_gaia_2021} \emph{Gaia} DR3 catalogue of globular clusters and the \cite{bica_general_2008} catalogue of star clusters in the Magellanic clouds. Names between catalogues were standardised as much as possible to facilitate easier comparison and remove duplicated clusters. One such example are ESO clusters, which are numbered based on their position in the form `ESO~XXX-XX' in the original work and \cite{kharchenko_global_2013}, but with numbers that are separated by a space instead of a dash in \cite{cantat-gaudin_clusters_2020} and \cite{dias_new_2002}, or often miss leading zeroes in \cite{bica_multi-band_2018}.

\subsection{Recovery of clusters from prior works}

\begin{table}
\caption{Results of crossmatching against literature catalogues sorted by $n_\text{clusters}$.}
\centering
\label{tab:crossmatches}
\begin{tabular}{l c c c}
\hline\hline
Work & $n_{\text{clusters}}$ & $n_{\text{detected}}$ & \% \\
\hline

\cite{bica_multi-band_2018} & 4391 & 1251 & 28.5 \\
\cite{kharchenko_global_2013} & 2935 & 1513 & 51.6 \\
\cite{dias_new_2002} & 2161 & 1160 & 53.7 \\
\cite{he_unveiling_hidden_2022} & 1656 & 737 & 44.5 \\
\cite{cantat-gaudin_clusters_2020} & 1481 & 1431 & 96.6 \\
\cite{hao_newly_2022} & 704 & 501 & 71.2 \\
\cite{castro-ginard_hunting_2022} & 628 & 558 & 88.9 \\
\cite{castro-ginard_hunting_2020} & 582 & 519 & 89.2 \\
\cite{he_new_2022} & 541 & 440 & 81.3 \\
\cite{he_blind_allsky_2022} & 270 & 122 & 45.2 \\
\cite{sim_207_2019} & 208 & 180 & 86.5 \\
\cite{qin_hunting_2023} & 101 & 74 & 73.3 \\
\cite{chi_lisc_2023} & 82 & 18 & 22.0 \\
\cite{liu_catalog_2019}\tablefootmark{a} & 76 & 57 & 75.0 \\
\cite{he_catalogue_2021}\tablefootmark{b} & 74 & 69 & 93.2 \\
\cite{li_lisc_2022} & 64 & 44 & 72.1 \\
\cite{chi_identify_2022}\tablefootmark{b} & 46 & 11 & 23.9 \\
\cite{hunt_improving_2021} & 41 & 41 & 100.0 \\
\cite{li_lisc_2023} & 35 & 0 & 0.0 \\
\cite{ferreira_new_2021} & 34 & 32 & 94.1 \\
\cite{ferreira_discovery_2020} & 25 & 25 & 100.0 \\
\cite{casado_new_2021} & 20 & 15 & 75.0 \\
\cite{hao_sixteen_2020}\tablefootmark{b} & 16 & 5 & 31.3 \\
\cite{jaehnig_membership_2021} & 11 & 7 & 63.6 \\
\cite{santos-silva_canis_2021} & 5 & 4 & 80.0 \\
\cite{qin_discovery_2021}\tablefootmark{b} & 4 & 4 & 100.0 \\
\cite{ferreira_three_2019} & 3 & 0 & 0.0 \\
\cite{casado_discovery_2023} & 2 & 2 & 100.0 \\
\cite{anders_ngc_2022-1} & 1 & 1 & 100.0 \\
\cite{bastian_gaia_2019} & 1 & 1 & 100.0 \\
\cite{tian_discovery_2020} & 1 & 1 & 100.0 \\
\cite{zari_3d_2018}\tablefootmark{b} & 1 & 1 & 100.0 \\

\hline

\cite{kounkel_untangling_2020}\tablefootmark{c} & 8281 & 1498 & 18.1\% \\
\cite{bica_general_2008}\tablefootmark{d} & 3740 & 22 & 0.6\% \\
\cite{vasiliev_gaia_2021}\tablefootmark{e} & 170 & 134 & 78.8\% \\

\hline

\end{tabular}

\tablefoot{
32 catalogues of OCs are listed in the first section of the table, in addition to three catalogues at the bottom of other star clusters.
\tablefoottext{a}{Original work and this work uses the acronym `FoF' to name clusters, although others list with acronym `LP'.}
\tablefoottext{b}{Cluster(s) in these works were unnamed, and so cluster acronyms were adopted based on first letters of surnames of authors.}
\tablefoottext{c}{Catalogue of predominantly moving groups, although many are also open clusters.}
\tablefoottext{d}{Position-only catalogue of objects in the Magellanic clouds.}
\tablefoottext{e}{Catalogue of globular clusters.}
}

\end{table}

Table~\ref{tab:crossmatches} shows that this work has a high recovery rate of OCs from other works. As shown in Table~\ref{tab:crossmatches}, we recover 96.6\% of clusters from \cite{cantat-gaudin_clusters_2020}, higher than the 86.4\% of clusters recovered in Paper 1. Generally, clusters not recovered in Paper 1 were sparse, barely-visible overdensities in \emph{Gaia} DR2 which often now stand out strongly in \emph{Gaia} DR3, including clusters such as Berkeley~91 and Auner~1, which we now detect reliably at S/Ns of 9.7$\sigma$ and 12.5$\sigma$ respectively. The fact that only \cite{cantat-gaudin_clusters_2020} was able to detect these clusters in DR2 is likely due to a difference in methodology -- by starting with prior cluster positions, their search regions for these clusters are smaller and may help the clusters to stand out. However, the disadvantage of such an approach is that it may also introduce a handful of false positives, due to poor statistics inherent in such small search regions -- in Paper 1, we comment that a handful of clusters in \cite{cantat-gaudin_painting_2020} may not exist, which may be the case for some of the 3.4\% of clusters we are still not able to recover in \emph{Gaia} DR3 despite the greatly improved astrometry and clear benefits to the S/N of other previously undetected clusters. 

We recover most of the new clusters reported in \cite{castro-ginard_hunting_2020} (a work based on \emph{Gaia} DR2) and \cite{castro-ginard_hunting_2022} (a work based on \emph{Gaia} EDR3), recovering almost exactly 89\% of both catalogues, showing that a majority of these objects can be confirmed independently. The reason for the non-recovery of around 11\% of clusters in both cases is not clear, although the fact that this amount is similar between both clusters detected with \emph{Gaia} DR2 and EDR3 suggests that it is a fundamental methodological difference (their works use the DBSCAN algorithm, see Paper 1 for a review) rather than a data one. 

However, we recover fewer of the new clusters reported by other DBSCAN-based works such as \cite{hao_sixteen_2020, hao_newly_2022} and \cite{he_catalogue_2021, he_blind_allsky_2022, he_new_2022, he_unveiling_hidden_2022}, recovering fewer than 50\% of the clusters reported in \cite{he_blind_allsky_2022, he_unveiling_hidden_2022} using \emph{Gaia} EDR3 data.

Additionally, while a large fraction of clusters reported before \emph{Gaia} and catalogued in works such as \cite{dias_new_2002}, \cite{kharchenko_global_2013}, and \cite{bica_multi-band_2018} still do not appear in \emph{Gaia} DR3, we are able to reliably detect an additional 277 clusters from \cite{dias_new_2002}, 292 clusters from \cite{kharchenko_global_2013}, and 127 clusters from \cite{bica_multi-band_2018} that do not appear in the \emph{Gaia} DR2 catalogue of \cite{cantat-gaudin_clusters_2020} (excluding GCs in all cases, as the catalogue of \cite{cantat-gaudin_clusters_2020} does not contain them.)

Notably, we are unable to detect any of the high galactic latitude OCs that have been reported recently in \cite{li_lisc_2023}, despite the fact that OCs at such high latitudes should stand out clearly against the low number of field stars in the galactic halo. This echoes the results of \cite{cantat-gaudin_characterising_2018} and \cite{cantat-gaudin_clusters_2020}, who also find that high latitude OCs that have been reported in works such as \cite{schmeja_global_2014} are undetectable in \emph{Gaia} data.

We discuss possible reasons for the non-detection of many literature OCs further in Sect.~\ref{sec:discussion-undetected}.

Finally, it is worth commenting on our detections of moving groups, globular clusters, and Magellanic cloud objects. We are only able to detect 18.1\% of moving groups and clusters from the catalogue of \cite{kounkel_untangling_2020}, despite this work using the same algorithm (HDBSCAN). Many of the groups reported in \cite{kounkel_untangling_2020} have large on-sky extents that are larger than the fields used in this work. However, although 2276 of their 8281 clusters are compact enough to be easily detectable in our fields, we only recover 622 (27.3\%) of these compact groups, many of which correspond anyway to known nearby OCs. In Paper 1, we found that while HDBSCAN is the most sensitive clustering algorithm for application to \emph{Gaia} data, it also reports a large number of false positives without additional postprocessing to remove clusters based on their statistical significance. It may be that these clusters are false positives, although this should be investigated further in detail \citep[see e.g.][]{zucker_disconnecting_2022}.

The recovery of a large fraction of GCs in \cite{vasiliev_gaia_2021} shows that HDBSCAN can be used to effectively recover GCs. The non-recovered objects are mostly distant and heavily reddened GCs whose member stars can only be recovered with a prior position and distance to narrow the search region. Finally, while not a focus of this work, the recovery of 22 Magellanic cloud star clusters from \cite{bica_general_2008} shows that \emph{Gaia} data could be used to make limited inferences on existing Magellanic cloud clusters in a future work, although we do not appear to detect any new clusters in the Magellanic clouds as their distance is too high. 

\subsection{Assignment of names}\label{sec:crossmatching:names}

\begin{figure}[t]
   \centering
   \includegraphics[width=\columnwidth]{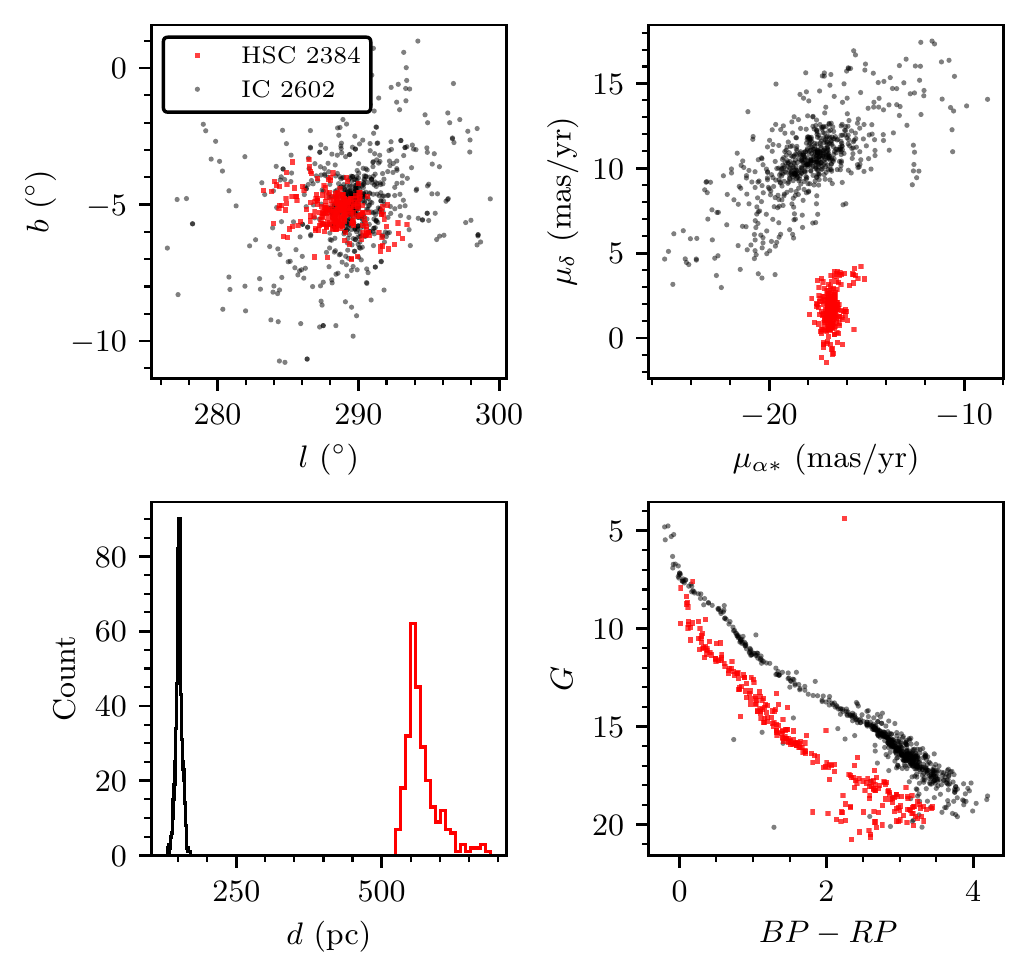}
   \caption{Member stars for the candidate new cluster HSC~2384 (red squares) compared against the nearby cluster IC~2602 (black circles). Four plots of are shown, comparing positions (top left), proper motions (top right) and photometry (bottom right). The bottom left plot shows a histogram of all distances to individual member stars.}%
   \label{fig:hsc_2384}
\end{figure}

\begin{table*}

\caption{Mean parameters for the clusters detected in this study.\label{tab:catalogue}}

\centering
\begin{tabular}{*{11}{c}}

\hline\hline

Name & ID\tablefootmark{a} & S/N & $n_\text{stars}$ & $\alpha$ ($^\circ$) & $\delta$ ($^\circ$) & $r_{50}$ ($^\circ$) & $\mu_{\alpha*}$ (mas yr\textsuperscript{-1}) & $\mu_{\delta}$ (mas yr\textsuperscript{-1}) & $\varpi$ (mas) & $\log t$ \\

\hline

\multicolumn{11}{c}{$\cdot \cdot \cdot$} \\ 
HSC 1 & 1805 & 8.21 & 64 & 289.61 & -38.03 & 3.32 & -1.029 (0.054) & -8.941 (0.085) & 2.097 (0.006) & 7.87$^{+0.24}_{-0.27}$ \\
HSC 2 & 1806 & 3.79 & 16 & 268.63 & -29.53 & 0.13 & 1.680 (0.031) & -1.182 (0.032) & 0.634 (0.003) & 7.92$^{+0.24}_{-0.22}$ \\
HSC 3 & 1807 & 3.89 & 24 & 273.73 & -31.87 & 0.12 & 0.371 (0.019) & 0.210 (0.025) & 0.647 (0.005) & 8.75$^{+0.18}_{-0.20}$ \\
HSC 4 & 1808 & 3.32 & 17 & 269.07 & -29.64 & 0.02 & 2.125 (0.067) & -11.895 (0.060) & 0.112 (0.015) & 7.54$^{+0.45}_{-0.50}$ \\
HSC 5 & 1809 & 4.38 & 18 & 276.78 & -33.09 & 0.12 & 0.150 (0.047) & -6.676 (0.049) & 0.657 (0.004) & 9.70$^{+0.30}_{-0.17}$ \\
HSC 6 & 1810 & 4.57 & 21 & 267.71 & -28.82 & 0.05 & -0.292 (0.017) & -1.516 (0.023) & 0.252 (0.004) & 7.84$^{+0.29}_{-0.27}$ \\
HSC 7 & 1811 & 3.12 & 18 & 261.40 & -25.13 & 0.09 & -5.033 (0.061) & -0.983 (0.060) & 0.464 (0.005) & 9.68$^{+0.32}_{-0.15}$ \\
HSC 8 & 1812 & 3.33 & 28 & 267.67 & -28.63 & 0.06 & 0.207 (0.014) & -0.211 (0.026) & 0.340 (0.004) & 7.86$^{+0.22}_{-0.23}$ \\
HSC 9 & 1813 & 5.88 & 25 & 269.05 & -29.33 & 0.16 & 2.120 (0.020) & -0.289 (0.021) & 0.549 (0.005) & 7.61$^{+0.22}_{-0.19}$ \\
HSC 10 & 1814 & 4.56 & 12 & 268.23 & -28.80 & 0.06 & -0.200 (0.011) & -1.753 (0.013) & 0.351 (0.003) & 8.20$^{+0.30}_{-0.31}$ \\
\multicolumn{11}{c}{$\cdot \cdot \cdot$} \\ 

\hline
\end{tabular}
\tablefoot{Standard errors for mean proper motions and parallaxes are shown in the brackets. The full version of this table with 7167 rows and many extra columns is available at the CDS only, with a complete description of the included additional data in Appendix~\ref{app:tables}.
\tablefoottext{a}{Internal designation used to link final catalogue entries to their crossmatching results in Table~\ref{app:tab:all_crossmatches}.}
}
\end{table*}

\begin{figure*}[t]
   \centering
   \includegraphics{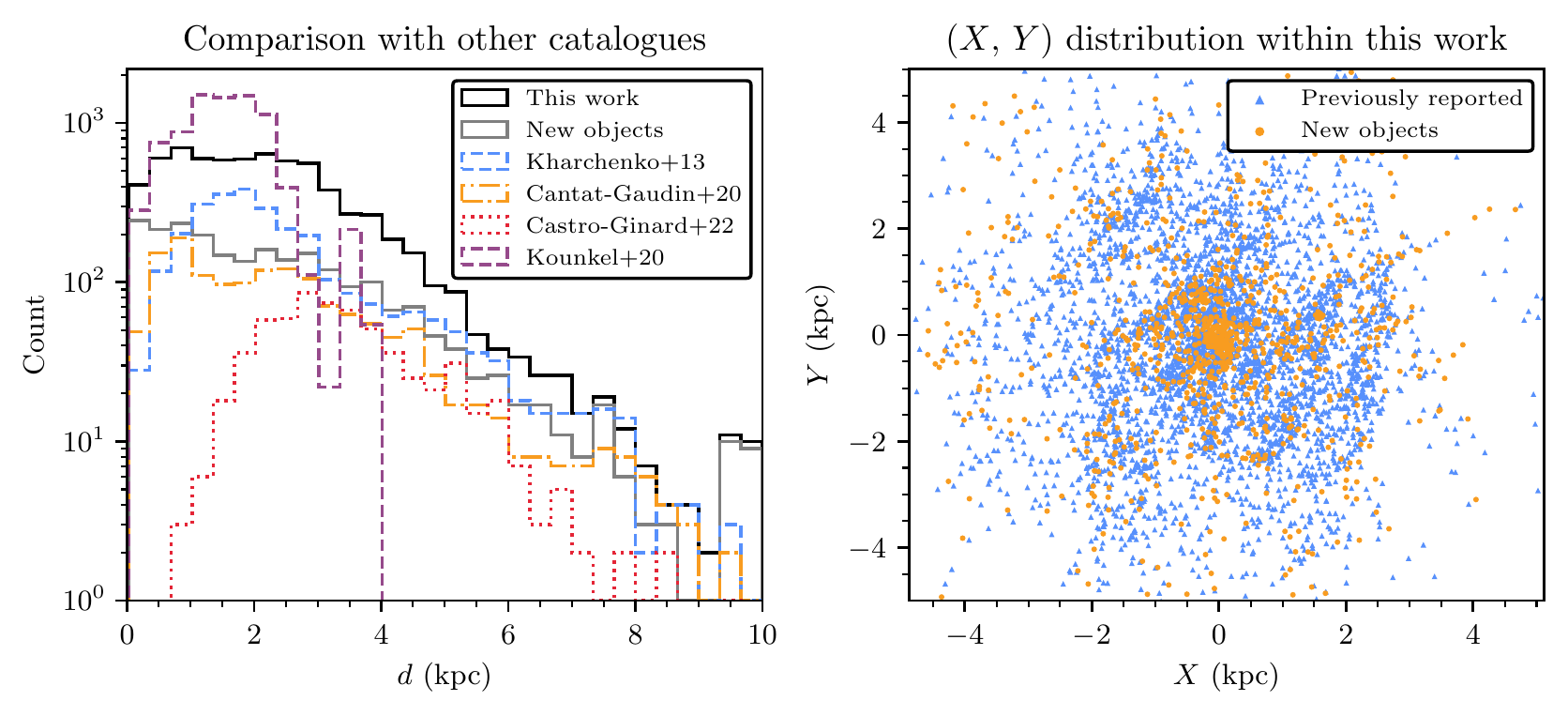}
   \caption{Distance and spatial distributions of clusters in this work. \emph{Left}: the distance distribution of all clusters in this work that do not crossmatch to known GCs compared to other catalogues. \emph{Right}: The distribution of clusters in this work in Cartesian coordinates centred on the Sun, cut to only those within 5~kpc in the X or Y directions. All previously reported clusters that we redetect are shown as blue triangles, and all objects new in this work shown as orange circles.}%
   \label{fig:distance_distribution}
\end{figure*}

As many of the objects we detect crossmatch to multiple entries in the literature (or vice-versa), assigning detected clusters to literature names can be non-trivial. A total of 7022 literature clusters crossmatch to 4944 of the entries in our catalogue, of which only 2749 matches are direct one-to-one matches where a single detected cluster can be easily assigned a single name.

1396 detected clusters each match to multiple literature entries. In these cases, the main cluster name was assigned based on the date of submission to a journal, with other names recorded in a separate column of alternative names for this object.

In 64 cases, multiple detected clusters crossmatched to the same literature object. The best match was selected based on position (or proper motions and distances, if available), with other objects instead recorded as new clusters.

Finally, there were 265 groups of crossmatches where multiple detected clusters crossmatched to multiple literature clusters, where assigning one match affects other matches. This is common in regions where many clusters are in a small area, such as in star formation regions like the Carina nebula. For simplicity, and since many of these groups contain literature entries with only positions available, we assign the best match on cluster positions only, iterating over all matches within a group accepting the match with the smallest positional separation and then removing all other literature entries with the same name within this group. All valid matches for every cluster are recorded in a separate column, and as these crossmatches represent the most difficult to assign reliably, clusters where their name has been assigned in this way are flagged in the catalogue as crossmatches that were particularly difficult to assign.

After assigning names to clusters, removing 22 objects associated with the Magellanic clouds, 17 objects associated with galaxies or dwarf galaxies, and 582 objects clearly associated with stellar streams in the galactic halo, our catalogue contains 7167 clusters, and is listed in Table~\ref{tab:catalogue} and online at the CDS, with tables of member stars and the rejected Magellanic cloud objects, galaxies, and stellar streams available online only. 2387 of these clusters are unreported in the literature and are candidate new objects, which we label with the acronym `HSC' (standing for HDBSCAN Star Cluster.) Most of these objects have good-quality CMDs, and some are likely to be new OCs. For instance, HSC~2384 is a nearby new OC candidate at a distance of only 551~pc with 273 member stars and a high astrometric S/N of $23.6\sigma$, which likely avoided prior detection due to being obscured by IC~2602 and mis-crossmatched to it (shown in Fig.~\ref{fig:hsc_2384}.) However, many appear to be more consistent with unbound moving groups, and will require further classification based on their structure and dynamics. In addition, we provide a table of all crossmatches and non-crossmatches against the clusters in this work in Table~\ref{app:tab:all_crossmatches}.

In the next sections, we discuss multiple aspects of the overall catalogue. Firstly, we discuss the overall catalogue of existing clusters in Sect.~\ref{sec:results-overall}, including its distribution and the quality of its membership lists. Section~\ref{sec:discussion-undetected} discusses why some literature clusters are undetected. Finally, Sect.~\ref{sec:discussion-moving_groups} discusses why existing approaches to differentiate between moving groups and OCs are inadequate to classify the new clusters detected in this work, a topic that will be explored further in a future work (Hunt~\&~Reffert,~\emph{in prep.}).

\begin{figure*}[t]
   \centering
   \includegraphics[width=\textwidth]{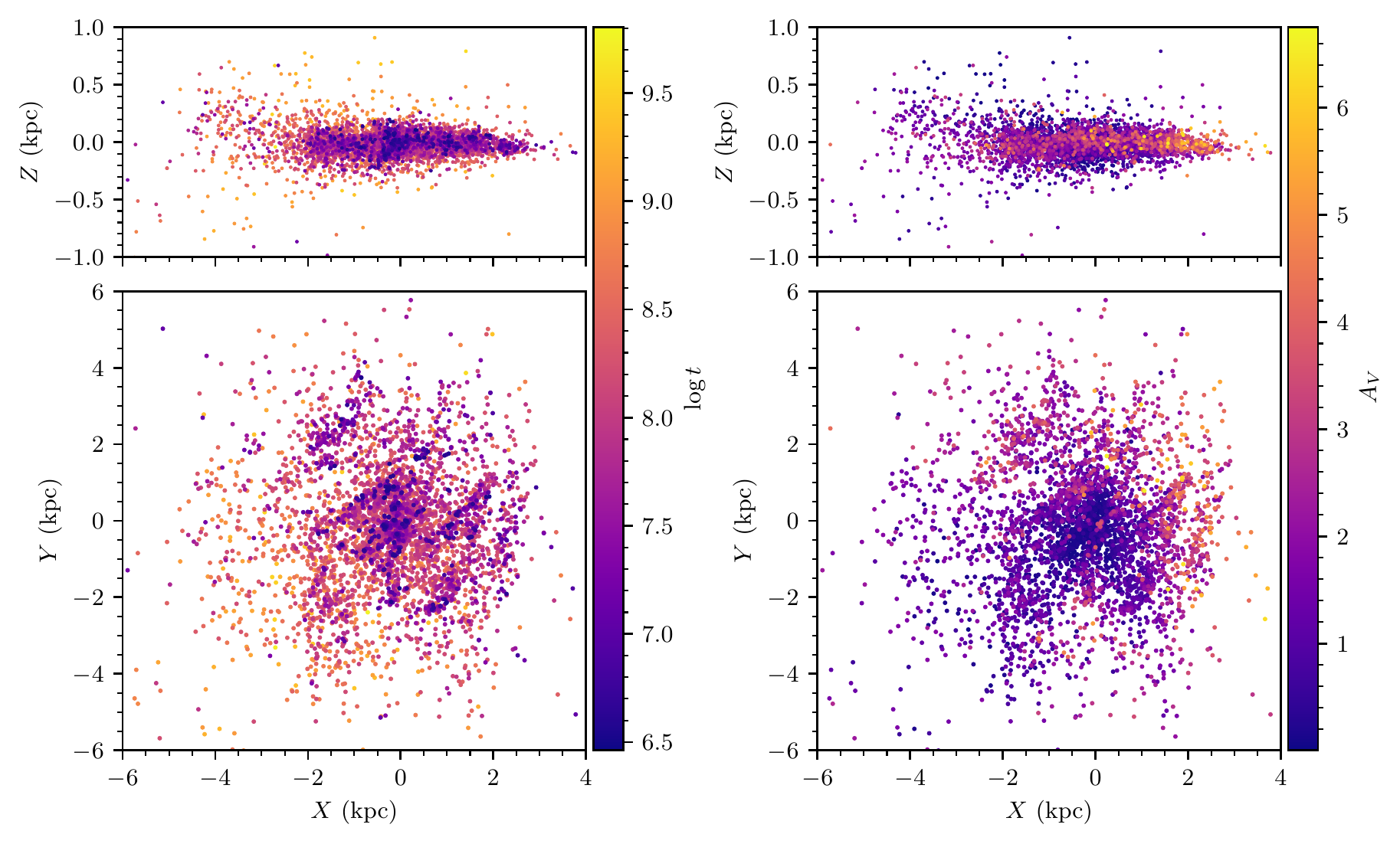}
   \caption{Spatial distributions of clusters detected in this work shaded on our derived $\log t$ and $A_V$ values. \emph{Left:} side-on and top-down distribution of clusters in heliocentric coordinates that do not crossmatch to known GCs. The galactic centre is to the right, with the Sun at $(0,0)$. Only clusters passing two quality cuts are plotted: firstly, those with a CST score above $5\sigma$, meaning they are highly probable astrometric overdensities; and secondly, a median CMD class above 0.5, which are those compatible with single population star clusters. Clusters are plotted in descending age order, meaning points representing young clusters are most visible in crowded regions. \emph{Right:} as left, except clusters are colour-coded by extinction $A_V$. Clusters are plotted in ascending order of extinction.}%
   \label{fig:age_av_distribution}
\end{figure*}

\begin{figure}[t]
   \centering
   \includegraphics{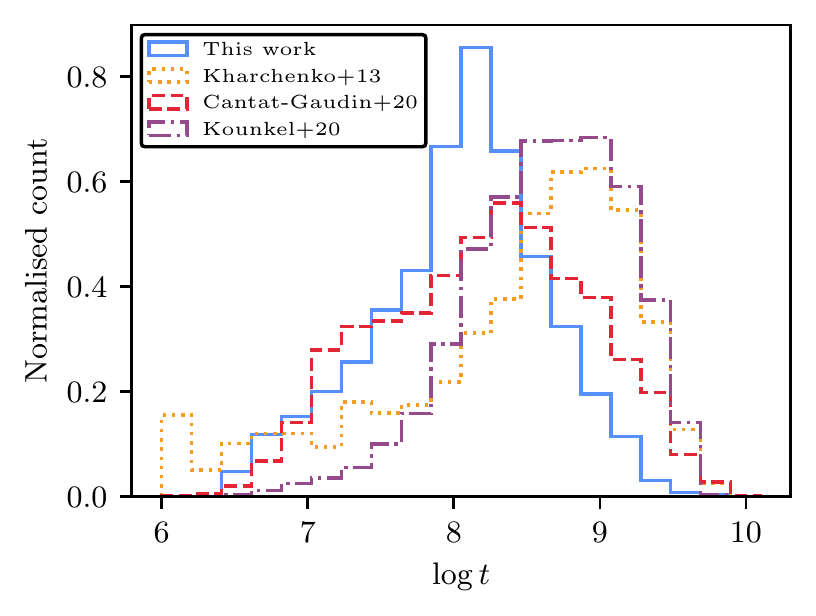}
   \caption{Histogram of ages of all clusters in this work with median CMD classes greater than 0.5 -- specifically, all clusters with photometry that is compatible with a single population of stars. These are compared to the ages of all clusters in the catalogues of \cite{kharchenko_global_2013}, \cite{kounkel_untangling_2020}, and \cite{cantat-gaudin_painting_2020}. Known GCs are excluded from the results of this work and the results of previous works for this plot.}%
   \label{fig:age_histogram}
\end{figure}

\begin{figure}[t]
   \centering
   \includegraphics{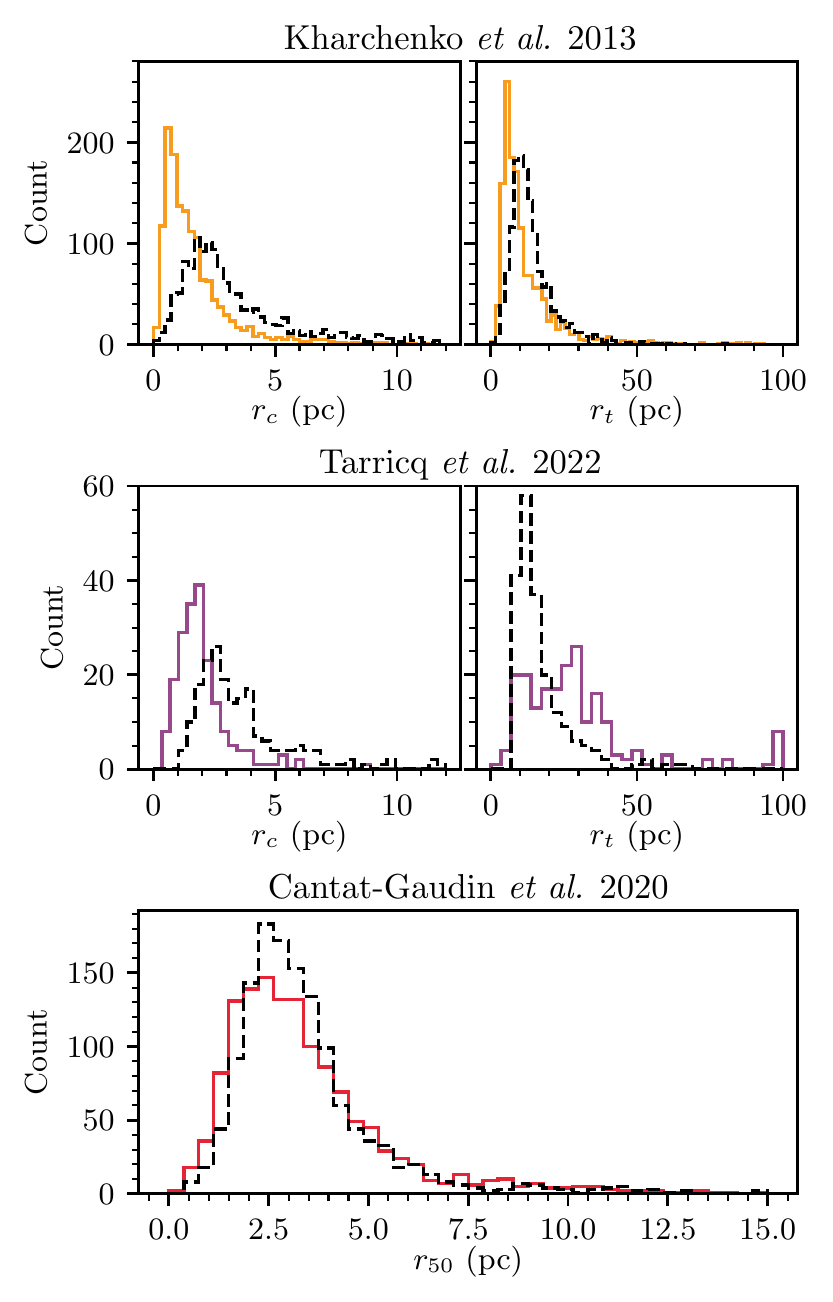}
   \caption{Cluster radii derived in this work (dashed black line) compared against the distributions of cluster radii in various literature works. \emph{Top row:} $r_c$ (top left) and $r_t$ (top right) of 1446 clusters from \cite{kharchenko_global_2013} that we redetect in this work (solid orange curve) compared against our approximately estimated \cite{king_structure_1962} radii for these 1446 clusters. \emph{Middle row:} same as top, except for radii of 202 clusters from \cite{tarricq_structural_2022} that have derived King radii (solid purple curve). \emph{Bottom:} $r_{50}$ measurements from \cite{cantat-gaudin_clusters_2020} compared against our $r_{50}$ measurements for the 1343 clusters from their work that we redetect.}%
   \label{fig:radius_distributions}
\end{figure}

\section{Overall results}\label{sec:results-overall}

In this section, we briefly discuss the structure and characteristics of the overall catalogue of 7167 clusters. 

\subsection{Suggested cuts on the catalogue for a high-quality cluster sample}\label{sec:results-overall:suggested-cuts}

Our catalogue also includes objects that we detect with CST scores as low as $3\sigma$, and objects with low-quality CMDs given the results of our classifier in Sect.~\ref{sec:cmd_classifier}. Such clusters are included in our catalogue for completeness, as a low-quality CMD may be caused by a poor detection of a real OC by our cluster recovery method, and a cluster with a low CST that is not a guaranteed astrometric overdensity may still be a real cluster that could be validated by a future \emph{Gaia} data release. However, these clusters are not particularly scientifically useful for studies of star clusters, as they cannot be validated as real within this work, or even with any currently available data. 

Hence, in discussions of the overall structure of our results, we predominantly discuss the most reliable sample of 4105 clusters within the catalogue: those with a median CMD class greater than 0.5, meaning that they are likely to be a largely homogeneous single population of stars as in OCs and moving groups, allowing some tolerance for blue stragglers and extended main-sequence turnoffs; and a CST of greater than $5\sigma$, corresponding to  clusters with a high likelihood of being real overdensities within \emph{Gaia} data and not simply a statistical fluctuation. The more tenuous 3062 objects excluded by this cut may still be used in some analyses, although with the caveat that these objects are less likely to be real star clusters.



\subsection{General distribution}

The distribution of clusters in our catalogue is generally similar to that of other \emph{Gaia}-based works such as \cite{cantat-gaudin_clusters_2020}, albeit with more stark differences when compared to those compiled before \emph{Gaia}, such as \cite{kharchenko_global_2013}. Comparisons are also useful to the catalogue of structures, moving groups, and star clusters of \cite{kounkel_untangling_2020} and papers based on \emph{Gaia} DR3 data that report new clusters, such as \cite{castro-ginard_hunting_2022}.

Figure~\ref{fig:distance_distribution} shows the distance distribution of clusters in this work, as well as the $X,Y$ distribution of clusters we re-detect and objects new to this work. Owing to the improved astrometry of \emph{Gaia} DR3 and the clustering method we use (see Paper~1), our catalogue has a high total number of clusters in most distance bins relative to other catalogues. As expected from the results in Paper~1, HDBSCAN is a cluster recovery technique sensitive across all distance ranges. However, HDBSCAN is sensitive to all clusters within \emph{Gaia} data, as it is unbiased on the shape of clusters it reports; hence, the catalogue contains a large number of moving groups, which are generally detected near to the Sun. The catalogue contains around 8x as many objects as the open cluster catalogue of \cite{cantat-gaudin_clusters_2020} within 500~pc, clearly visible as an overdensity of new objects and in the distance distribution of Fig.~\ref{fig:distance_distribution}. These objects are often difficult to classify as being OCs or moving groups (see Sect.~\ref{sec:discussion-moving_groups}).

The age and extinction distribution of Fig.~\ref{fig:age_av_distribution} is similar to that of \cite{cantat-gaudin_painting_2020}. A number of structures stand out, including: the imprint of the galactic warp in $X,Z$ plots for $X<-2$~kpc; the presence of spiral arm structure amongst young clusters very similar to that reported in works such as \cite{castro-ginard_milky_2021}; and the general flatness of the distribution of compact star clusters in the Milky Way other than GCs, with few existing at heights of $\lvert Z \rvert > 250$~pc. Additionally, clusters towards the galactic centre generally have high $A_V$ values of 5 or greater, suggesting that extinction may be a limiting factor in the detection of clusters in this direction.

Differences to pre-\emph{Gaia} works are most apparent in the age histogram of Fig.~\ref{fig:age_histogram}, however. Our combined age distribution is relatively similar to that of \cite{cantat-gaudin_painting_2020}, albeit with a slightly lower median age around $\log t \approx 8$ and no additional bump between $7 < \log t < 8$. However, the star cluster catalogue of \cite{kharchenko_global_2013} skews significantly older, with the most common (modal) age for clusters being around $\log t \approx 9$, an age range where we detect few clusters. A similar pattern is also visible for the catalogue of \cite{kounkel_untangling_2020}, whose moving group and star cluster catalogue contains many unbound, old structures. Many of these objects have similar ages to the typical ages of unclustered stars in the Milky Way disk. In Sect.~\ref{sec:discussion-undetected}, we elaborate on how some of these age differences may be caused by these catalogues containing a number of old false positive clusters. 

Finally, Fig.~\ref{fig:radius_distributions} shows the distribution of cluster radii compared between this work and the works of \cite{kharchenko_global_2013}, \cite{tarricq_structural_2022}, and \cite{cantat-gaudin_clusters_2020}. Our cluster radii agree most strongly with those in \cite{cantat-gaudin_clusters_2020}, with a similar distribution of cluster radii containing 50\% of members $r_{50}$. The \cite{king_structure_1962} core radii $r_c$ that we derive, when compared against those in \cite{kharchenko_global_2013} and \cite{tarricq_structural_2022}, are generally larger. This may be due to our more populated membership lists, particularly for faint stars, due to our lack of a magnitude cut in our clustering analysis. Particularly for clusters with a high degree of mass segregation, this difference in memberships would cause our clusters to have larger observed cores. Our tidal radii $r_t$ are slightly larger than those in \cite{kharchenko_global_2013}, but much smaller than those in \cite{tarricq_structural_2022}. In the first case, the difference may be due to the improved precision of \emph{Gaia} data compared to pre-\emph{Gaia} works, causing us to detect more member stars at the outskirts of clusters and hence derive larger cluster tidal radii, with this effect again being stronger for mass segregated clusters. In the second case, since \cite{tarricq_structural_2022} also explicitly searched for cluster tidal tails and comas in their work, it may be that their extended cluster membership lists mean that they report higher cluster tidal radii.


\begin{figure*}[t]
   \centering
   \includegraphics{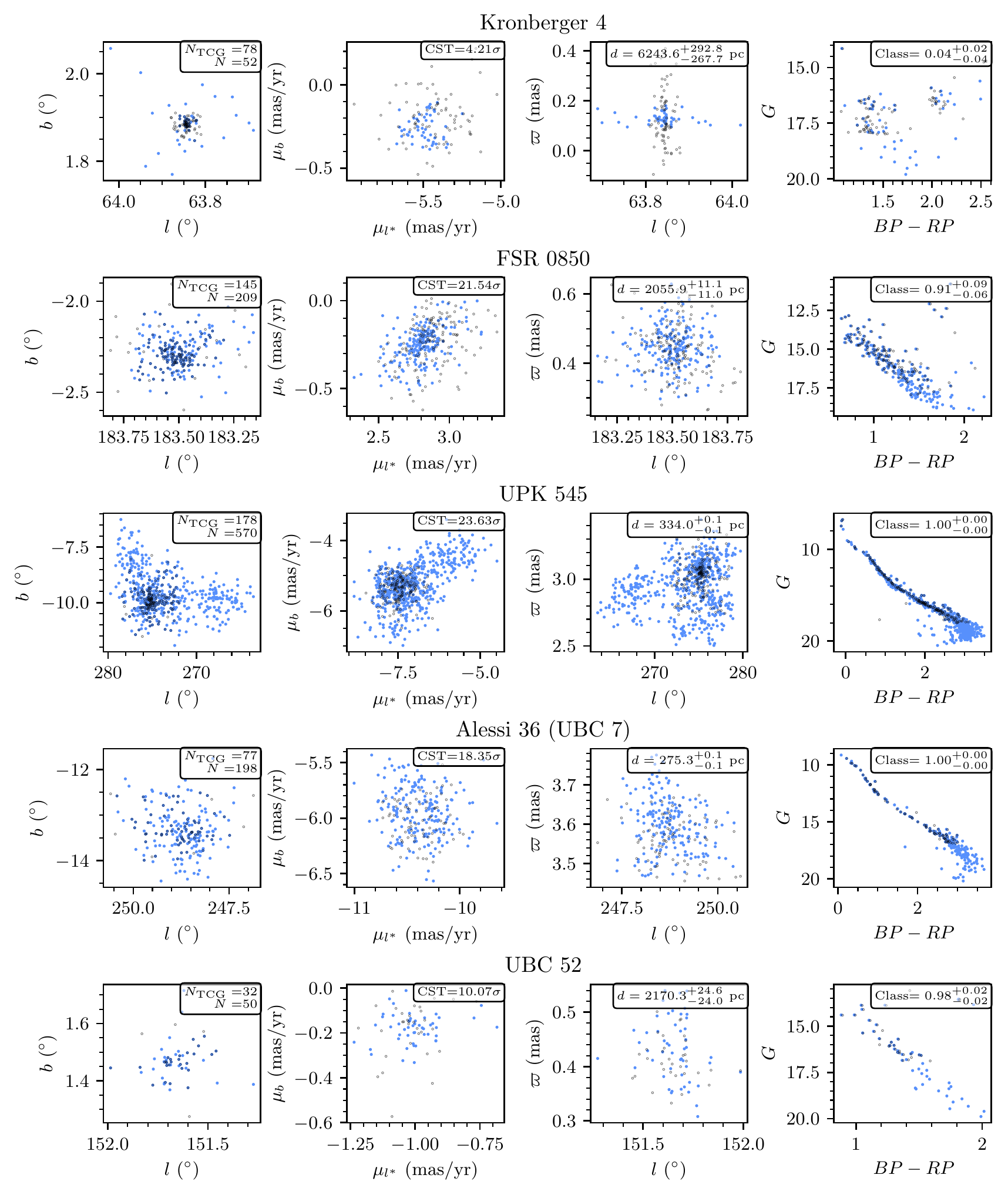}
   \caption{Membership list comparisons between this work and the catalogue of \cite{cantat-gaudin_clusters_2020}, using three clusters selected at random (upper three) and two clusters selected at random that were detected in \cite{castro-ginard_new_2018} using \emph{Gaia} DR1 data. Stars assigned as members by this work are plotted with filled blue circles, while members reported by \cite{cantat-gaudin_clusters_2020} are plotted with empty black circles. The first three columns compare the astrometry of cluster members in galactic coordinates, proper motions, and parallax as a function of $l$. The final column compares colour-magnitude diagrams of each resulting membership list. For every cluster, various parameters are labelled on the plots: number of member stars in \cite{cantat-gaudin_clusters_2020} $N_{\text{TCG}}$, number of member stars in this work $N$, astrometric S/N as estimated by the CST, distance $d$, and probability of being a single stellar population given the neural network in Sect.~\ref{sec:cmd_classifier}.
   }%
   \label{fig:membership_comparison}
\end{figure*}

\clearpage

\subsection{Membership lists for individual clusters}

\begin{figure}[t]
   \centering
   \includegraphics[width=\columnwidth]{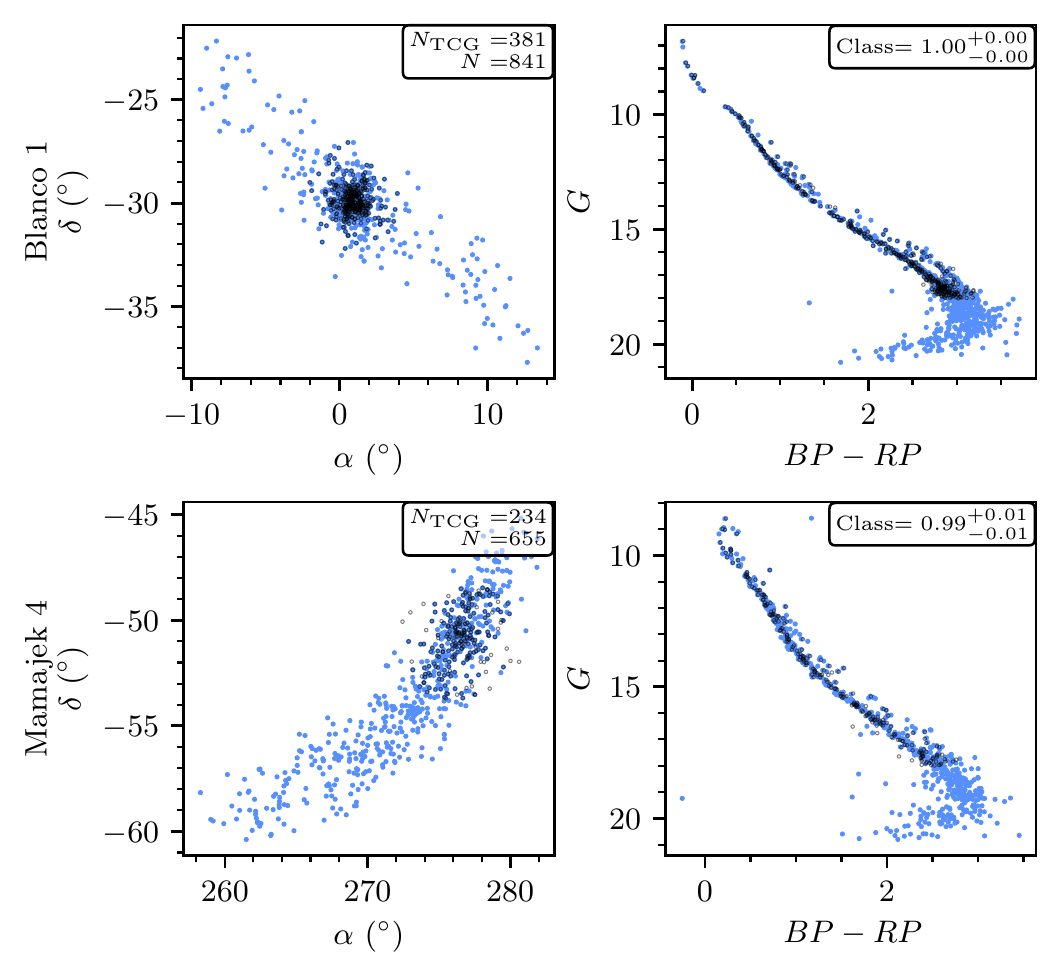}
   \caption{Two examples of clusters in the catalogue that have detected tidal structures. The spatial distribution of the clusters Blanco~1 (top row) and Mamajek~4 (bottom row) are plotted on the left, with member stars reported in this work shown as filled blue circles and compared against member stars from \cite{cantat-gaudin_clusters_2020} which are plotted as empty black circles. CMDs are shown in the two plots on the right for both clusters.}%
   \label{fig:example_cluster_ra_dec}
\end{figure}

Owing to the improved quality of \emph{Gaia} DR3 data and the expanded selection of 729~million stars from \emph{Gaia} data used as input into our cluster recovery pipeline, clusters in this work generally have more populated membership lists than in previous catalogues. Fig.~\ref{fig:membership_comparison} compares our membership lists with those from \cite{cantat-gaudin_clusters_2020} for five clusters randomly selected from our catalogue. Our membership lists typically have a higher total number of stars, with virtually all new member stars being compatible with the existing cluster CMD. This is particularly the case for clusters in regions with minimal crowding, where \emph{Gaia} has a high completeness of stars with 5-parameter astrometry down to $G\sim20$, with our membership lists containing stars down to approximately this limit. For more distant clusters such as Kronberger~4, membership lists are comparable in quality to those of \cite{cantat-gaudin_clusters_2020}, as \emph{Gaia} DR3 data does not present a large improvement in the astrometric quality of these distant sources compared to DR2. On average, our work contains $2.1$ times as many member stars as the clusters we have in common with \cite{cantat-gaudin_clusters_2020}, and $4.1$ times as many member stars as the clusters we have in common with \cite{kharchenko_global_2013}.

A second major advantage of our pipeline is that clusters are not forced to take a spherical shape, as with other methods such as Gaussian mixture models (Paper 1). Hence, we are able to detect tidal tails for many of the clusters in the catalogue, especially for those that are nearby and within $1-2$~kpc. \cite{tarricq_structural_2022} use HDBSCAN to detect tidal tails for 71 nearby OCs, many of which we are also able to detect. Figure~\ref{fig:example_cluster_ra_dec} shows two examples of nearby clusters with well-resolved tidal tails using our methodology, Blanco~1 and Mamajek~4, both of which have reported tidal tails stretching around 50~pc from the centre of the cluster. Virtually all stars within the tidal structures appear compatible with the isochrone of the cluster core, suggesting that they are stars with the same age, composition, and origin as the stars in the cluster cores. Particularly for clusters within 1~kpc, many of the clusters in our catalogue have tidal tails or comas.

However, as no current methodology for star cluster recovery from \emph{Gaia} data is perfect (Paper~1), our membership lists are not without caveats -- both of which are consistent with our results from Paper~1, but that are still worth mentioning in the main work of this catalogue. 

Firstly, for distant OCs, our method may return fewer members than some other approaches. At high distances ($d \gtrsim 5$~kpc), the errors on \emph{Gaia} parallaxes and proper motions generally become much higher than the intrinsic dispersion of OCs, meaning that many members have low membership probabilities and can only be reliably assigned as members by incorporating error information. Our methodology does not use error information in the clustering analysis for reasons of speed and the fact that HDBSCAN does not directly include a way to consider errors on data in clustering analysis, although other methods such as UPMASK \citep{krone-martins_upmask:_2014} which do consider error information could return better membership lists for these distant clusters. This is visible for Kronberger~4 in Fig.~\ref{fig:membership_comparison}, where the membership list of \cite{cantat-gaudin_clusters_2020} (which was compiled using UPMASK) has a slightly higher number of sources than our membership list, even though our list was compiled from a greater number of input sources due to our lack of a $G$-magnitude cut.

Secondly, HDBSCAN may sometimes return too many members, selecting regions larger than just an OC's core and tidal tails. This is particularly common for young clusters, which are often embedded in regions of high stellar density where recent hierarchical star formation has occurred \citep{portegies_zwart_young_2010}. These clusters can be difficult for HDBSCAN to isolate from other surrounding stars and sub-clusters. One particular example can be seen for UPK~545 in Fig.~\ref{fig:membership_comparison}. Although the tail emerging from the cluster core in the upper-left of the $(l,b)$ plot appears compatible with a tidal tail, the connected structure to the right of the cluster is not. It appears to have the same age and composition as the cluster core, with all members of the tail being photometrically consistent with it. However, this `offshoot' from the cluster may be better described as a separate cluster, which may also be bound to the core of UPK~545 in a binary pair of clusters, due to their proximity. Edge cases such as these are impossible to deal with autonomously with our current methodology and HDBSCAN alone, and require manual selection and separation of certain clusters in the catalogue into multiple separate components.

On a whole, the primary advantage of our catalogue is its completeness, generally reporting more member stars than previous works in the literature and doing so with a homogeneous methodology for a high number of total clusters. However, this is also the primary disadvantage of our catalogue: there are too many clusters and too many edge cases for all membership lists to be perfect, given only one clustering methodology. Hence, users of the catalogue who work with a small enough number of clusters are encouraged to manually check cluster membership lists and refine them depending on their application. To give one example, a user who wishes to only study cluster cores could refine our cluster membership lists by selecting a subset of them with Gaussian mixture models. With careful manual tweaking of the parameters of the mixture models, such a method could be used to remove tidal tails or possible other cluster components from our membership lists where necessary. Having discussed the general results of clusters in our catalogue, we next discuss the reasons why many clusters reported in the literature may not appear in our catalogue.


\section{Reasons for the non-detection of some literature objects}\label{sec:discussion-undetected}

\begin{figure*}[t]
   \centering
   \includegraphics[width=0.95\textwidth]{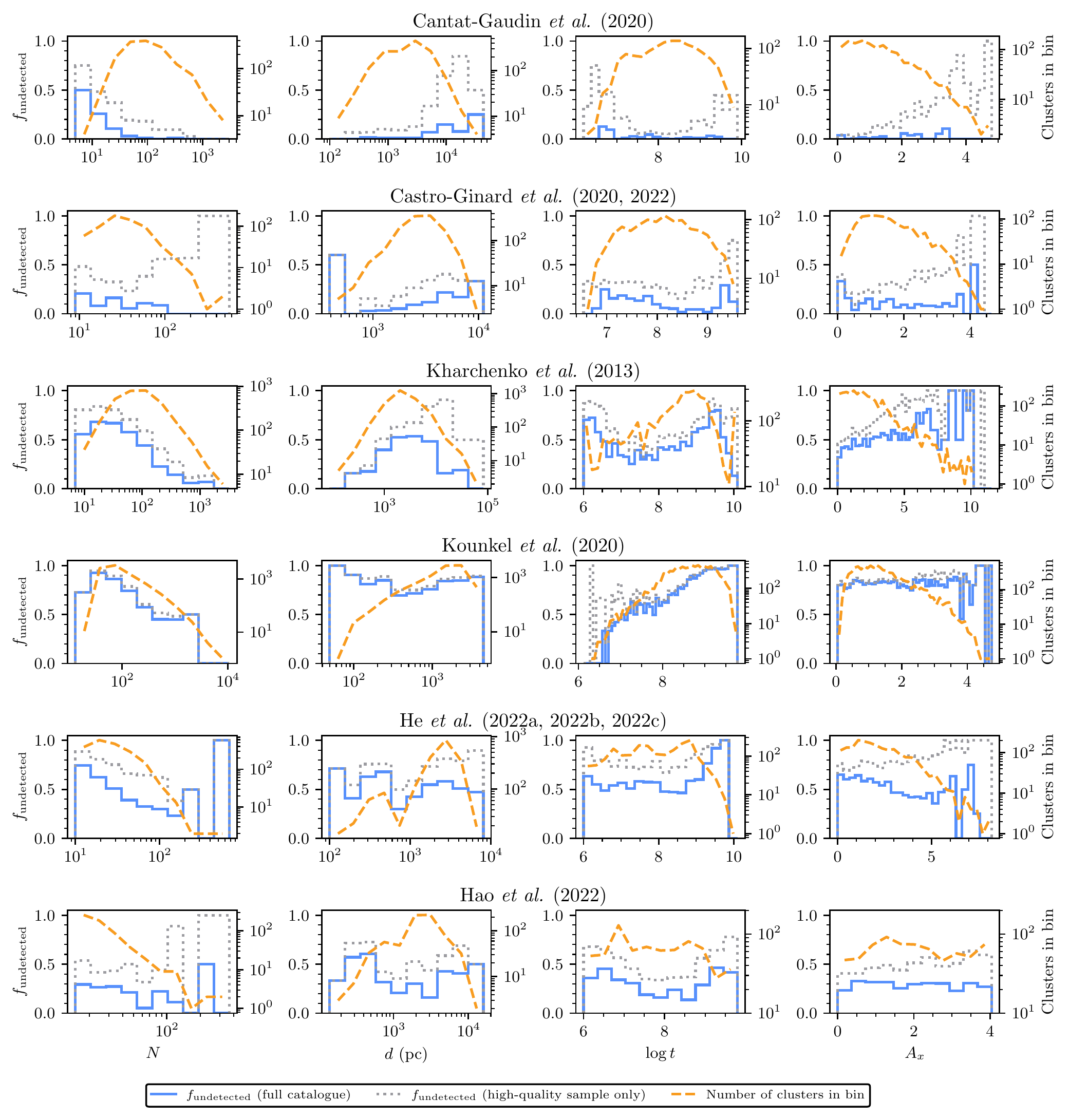}
   \caption{Plots showing the fraction of clusters undetected by this work when compared to various literature works or series of literature works, shown as a histogram of various parameters as a solid blue line for all clusters in the catalogue, and a dashed grey line for clusters in the high quality sample defined in Sect.~\ref{sec:results-overall:suggested-cuts}. The dashed orange lines show the number of clusters in each bin. Optimum histogram bin widths were selected automatically using \texttt{numpy} \citep{harris_array_2020}. From left to right, each column shows the number of stars $N$, distance $d$, age $\log t$ and extinction $A_x$ reported in each catalogue. For the top four groups of catalogues, extinctions were given in the $V$ band. For the lower two, extinctions were given in \emph{Gaia's} $G$ band, which are generally slightly lower.}%
   \label{fig:crossmatch_comparison}
\end{figure*}

Thousands of new OCs and moving groups have been reported since the release of \emph{Gaia} DR2 \citep{brown_gaia_2018}, with over 2000 reported in the last two years using \emph{Gaia} DR3 data alone \citep{gaia_collaboration_gaia_2021}. While multiple works have commented on the reliability of individual clusters in the literature at-length \citep[e.g.][]{cantat-gaudin_clusters_2020, piatti_assessing_2023}, as an unbiased search for all clusters within all of \emph{Gaia} DR3, the results of this work offer a unique way to review the reliability of recently detected OCs on a large scale. In addition, with hundreds of literature OCs newly redetected in this work, this work also offers a chance to update the status of many older clusters reported in the pre-\emph{Gaia} era.

The non-detection of a cluster by this work can be a result of multiple different factors. It is important to first rule out any possible methodological reasons before claiming that a given cluster does not exist. In Paper~1, we showed that our methodology has a high sensitivity, and hence a literature cluster being non-detected in this work can nevertheless raise strong doubts about whether or not it is real. With thousands of non-detected clusters, there are far too many to review all clusters individually, and hence we do not aim to decisively prove that some literature clusters are not real. We discuss the six main methodological and data-related reasons why a cluster may not appear in this work, concluding with questioning the existence of many objects reported in existing literature works.


\subsection{Methodological reasons for the non-detection of a cluster}\label{sec:discussion-undetected:methodological-reasons}
\subsubsection{Limitations of the clustering algorithm used}\label{sec:discussion-undetected:methodological-reasons:1-algorithm}

An obvious reason why we may not detect a given literature OC is due to limitations of the HDBSCAN algorithm that we use in this work. While we found in Paper~1 that HDBSCAN is the most sensitive clustering algorithm overall, DBSCAN was slightly more sensitive for clusters at distances greater than 5~kpc when applied to \emph{Gaia} DR2 data. On the other hand, with respect to cluster size, HDBSCAN was the most sensitive algorithm for all sizes of cluster, although HDBSCAN and DBSCAN had similar or identical sensitivity for clusters with a number of members stars of $n_\text{stars}=10$. Age and extinction were not found to have any significant differential impact on the sensitivity of the algorithms trialed, with all algorithms being more or less equally affected by older and/or heavily reddened clusters having fewer visible member stars, and hence being harder to detect.

The main limitation of HDBSCAN should be for clusters at distances greater than 5~kpc. However, only 6\% and 21\% of clusters from the DBSCAN-based works of \cite{castro-ginard_hunting_2020} and \cite{castro-ginard_hunting_2022} respectively that we are unable to detect have reported parallaxes of less than 0.2~mas, suggesting that distance-related detection issues alone are not enough to explain why certain clusters from these works are not detected. Additionally, we note that \cite{castro-ginard_hunting_2022} using \emph{Gaia} EDR3 were only able to recover $\gtrsim$80\% of clusters they found in DR2 in \cite{castro-ginard_hunting_2020}, and so DBSCAN itself between \emph{Gaia} data releases is not able to reliably reconfirm all clusters it detected previously.

Nevertheless, Fig.~\ref{fig:crossmatch_comparison} shows that our chance of recovering clusters at high distances can be lower for certain works. In particular, although we are unable to recover only 3.4\% of clusters reported in \cite{cantat-gaudin_clusters_2020}, most of the clusters from their work that we are unable to recover are small clusters at distances above 5~kpc, suggesting that an algorithmic limitation may contribute to why we are unable to recover remaining objects from \cite{cantat-gaudin_clusters_2020}. A key difference between our work and \cite{cantat-gaudin_clusters_2020} is that their work used locations of clusters reported in the literature to narrow their search regions, which may in some cases be enough to make very distant clusters at the absolute limit of detectability in \emph{Gaia} stand out. Future \emph{Gaia} data releases with better data should provide additional clarity on whether or not such objects are real.

\subsubsection{Differences in the definition of an OC}\label{sec:discussion-undetected:methodological-reasons:2-definition}

There is no single agreed upon definition of an OC in the literature, and the slight differences in definition between works could cause some clusters to be detected or missed.

Principle amongst these definitions is the minimum number of observed member stars for a valid cluster, $n_\text{stars, min}$, which is important to distinguish star clusters from multiple star systems, also being used by some works as a proxy for the significance of a cluster relative to the field. In the literature, values of $n_\text{stars, min}$ range from 8 in \cite{castro-ginard_hunting_2022} to as high as 50 in \cite{liu_catalog_2019}, with most works coalescing around a value of between 10 and 12 \citep{krumholz_star_2019}. For the purposes of this work, we adopt a value of 10, and we should hence miss very few literature clusters due to this constraint alone.

Secondly, OCs generally have a population of stars with the same age and chemical composition, due to forming at the same time from the same molecular cloud \citep{cantat-gaudin_milky_2022}. In practice, this is a difficult definition to constrain observationally, with the CMDs of OCs being broadened by effects such as differential extinction or outliers which are not true member stars, with these effects being worse with increasing distance and field star density. In addition, many OCs are not perfect single populations, with some hosting blue stragglers or having a clear second population in the form of an extended main-sequence turnoff \citep{cantat-gaudin_milky_2022}. For the purposes of this work, we classify our clusters with our CMD classifier (see Sect.~\ref{sec:cmd_classifier}) and include all clusters in the final catalogue, instead leaving the task of removing clusters with poor photometry to the end user (recommending a minimum class value of 0.5). This means that no clusters are missing from the catalogue due to photometric reasons.

Finally, OCs must be distinguished from other types of single-population stellar overdensities. Star clusters can be divided into bound clusters (such as OCs and GCs) and unbound clusters (typically referred to as moving groups). Some works, such as \cite{cantat-gaudin_clusters_2020}, use basic cuts on mean parameters to remove clear moving groups from their catalogue; we leave the classification of moving groups in our catalogue to a future work (Hunt \& Reffert, \emph{in prep.}) for reasons discussed in Sect.~\ref{sec:discussion-moving_groups}, and hence, no OCs are be missing from this work due to being catalogued as moving groups. We do, however, flag known GCs in our catalogue by crossmatching against the catalogue of GCs of \cite{vasiliev_gaia_2021}, with GCs in the Milky Way being distinguished from OCs by their age, which is typically greater than $\sim 6 Gyr$, and their mass, which is typically greater than $\sim 10^4 M_\sun$, whereas most OCs have masses no higher than $\sim 5000 M_\sun$ \citep{kharchenko_global_2013}. In total, differences in the fundamental definition of an OC between works should have a small impact on the inclusion of OCs in this work when compared to others.

\subsubsection{Different quality cuts between different works}\label{sec:discussion-undetected:methodological-reasons:2-cuts}

Different works in the literature often place different quality cuts on their catalogues, meaning that another possible reason why a given literature cluster does not appear in this catalogue would be if it has been cut for quality reasons. Our catalogue adopts a philosophy of allowing users to decide their own quality cuts as much as possible, and hence includes all objects with bad photometry as well as moving groups that are unlikely to be bound OCs. The approach of allowing end users of the catalogue to define their own quality cuts is a similar philosophy to how \emph{Gaia} data releases include many poor-quality sources, instead allowing users decide how strongly they wish to cut the \emph{Gaia} catalogue \citep{gaia_collaboration_gaia_2021}. Poor photometry and the bound or unbound status hence do not impact our recoverability of clusters in Fig.~\ref{fig:crossmatch_comparison}.

However, the sole quality cut applied to the catalogue that would affect its sensitivity is a cut on the astrometric S/N of detected clusters (derived using the CST) at 3$\sigma$. This was performed because clusters with an S/N below this threshold are likely to be false positives, and because the high number of clusters below this threshold greatly complicated the process of merging results between different runs (see Sect.~\ref{sec:clustering}). Including such a quality cut dramatically improved the run merging process and hence our membership lists and completeness for reliable clusters, which is a more important scientific product than a list of low quality clusters that we cannot deem likely to be real clusters based on their S/N alone.

While we believe this is a fair trade-off to produce a catalogue that is as reliable as possible overall, it is likely that some real clusters are be missed due to this cut on S/N. For instance, in Paper~1 using \emph{Gaia} DR2 data, we tentatively detected Teutsch~156 with an S/N of 0.68$\sigma$, which counted as a non-detection; however, using \emph{Gaia} DR3, we clearly detect Teutsch~156 with an S/N of 16.3$\sigma$. It is difficult to know exactly how many real literature clusters are missed due to this cut, particularly since some clusters in the literature with an S/N below 3$\sigma$ are likely to be statistical fluctuations and not real clusters, especially for S/Ns below 1$\sigma$. This can be approximately estimated using the histogram of detected cluster S/Ns in Fig.~\ref{fig:cst_histogram}. Since the distribution of literature cluster S/Ns is roughly flat for S/Ns below 10$\sigma$, assuming that this trend continues for S/Ns below 3$\sigma$, we may have missed approximately $\sim 300$ crossmatches to clusters reported before \emph{Gaia} DR3 and an additional $\sim 400$ reported using \emph{Gaia} DR3 data -- although, owing to the low S/Ns that such objects would inevitably have, it is also likely that a number of these crossmatches would be false positives.

Inevitably, a repeat of this work with better data (such as \emph{Gaia} DR4) would likely detect more of the objects that we do not recover with a sufficient statistical significance using \emph{Gaia} DR3 data. In the future, further development of clustering algorithms that produce fewer false positives and can be ran on more data at once (both of which would tremendously simplify the run-merging process) would allow the minimum S/N threshold to be lowered.

\subsubsection{When two clusters are catalogued as one cluster}\label{sec:discussion-undetected:methodological-reasons:3-two-in-one}

Certain other non-detections can be explained by further methodological differences. Sometimes, clusters reported as multiples in the literature are reported as a single object by HDBSCAN, even across all of its $m_{clSize}$ runs. A notable example is UPK~533 from \cite{sim_207_2019}, which was re-detected by \cite{cantat-gaudin_clusters_2020}, but which HDBSCAN assigns as simply being a member of a tidal tail of a different and significantly larger nearby cluster, UPK~545, with no HDBSCAN $m_{clSize}$ run separating the two objects. UPK~545 is shown in Fig.~\ref{fig:membership_comparison} on the third row. In this and other edge cases, our catalogue merges the two objects. An improved clustering algorithm that can separate edge-case binary clusters such as these autonomously would be helpful. However, only a small fraction of clusters (fewer than 1\%) are affected by this issue.

\subsubsection{When a literature catalogue's parameters deviate too strongly from a detected cluster}\label{sec:discussion-undetected:methodological-reasons:4-literature-error}

While our crossmatching procedure as outlined in Sect.~\ref{sec:crossmatching} aims to be as fair as possible, generally giving the benefit of the doubt to potential crossmatches, there are nevertheless cases where clusters reported in the literature still remain outside of our bounds for an accepted match. Generally, in all cases where this occurs, our detected cluster is significantly different to the literature object in at least one of the parameters considered for crossmatching, with these clusters representing ambiguous cases where it is not clear that the reported literature cluster is truly the same object.

CWNU~528 as reported in \cite{he_new_2022} is one example of a cluster reported in the literature that we are unable to detect within our crossmatching criteria. CWNU~528 is reported in \cite{he_new_2022} with 24 member stars, but appears to be a small offshoot of the recently reported new cluster OCSN~82 from \cite{qin_hunting_2023}, which has an overall position different by around 3$^\circ$ and a total of 157 member stars. CWNU~528 is so much smaller than OCSN~82 and at such a different location that it does not crossmatch to it given our adopted crossmatching scheme, even though a few of the member stars in our detection of OCSN~82 are in common with CWNU~528 and they have similar proper motions and parallaxes.

This case is likely to have been repeated a few times, and appears particularly common with clusters detected in \emph{Gaia} data using the DBSCAN algorithm \citep[as in][]{he_new_2022}. In Paper 1, we commented that while DBSCAN has an excellent sensitivity and low false positive rate (depending strongly how the $\epsilon$ parameter is chosen), it often had the sparsest and most incomplete membership lists of all algorithms we studied. Hence, detections of clusters may be so different or poor compared to what another algorithm recovers that crossmatch criteria may not be fulfilled, even when using a very permissive crossmatching scheme. In these cases, it is debatable whether the literature cluster is even the same object as the newly detected one.


\subsubsection{Limitations of \emph{Gaia} data}\label{sec:discussion-undetected:methodological-reasons:5-gaia-limitations}

Finally, it is worth considering the limitations of \emph{Gaia} data itself, particularly when comparing our catalogue to works created from different data sources. Notably, the catalogue of \cite{kharchenko_global_2013} was compiled before \emph{Gaia} and used infrared data from 2MASS \citep{skrutskie_two_2006}. \cite{cantat-gaudin_clusters_2020} are unable to recover a majority of the clusters from \cite{kharchenko_global_2013} using \emph{Gaia} DR2 data, and we are unable to recover 48.4\% of the clusters reported in their catalogue in \emph{Gaia} DR3 data. Given that infrared light is significantly less affected by extinction than the visual light used to compile \emph{Gaia} data, it begs the question of whether many clusters from \cite{kharchenko_global_2013} may still be missing from \emph{Gaia}-based catalogues due to extinction limits.

However, Fig.~\ref{fig:crossmatch_comparison} shows that extinction does not appear to play a major role in the non-detection of many clusters from \cite{kharchenko_global_2013}. If extinction was a major contributor to why we are unable to detect so many of the clusters in their catalogue, then one would expect to see a linear trend in $f_\text{undetected}$; all of their low-extinction clusters would be easily detected in \emph{Gaia}, until some cut-off value beyond which \emph{Gaia} detects no further clusters. On the contrary, most of their clusters have $A_V < 5$, and we are unable to detect around 50\% of all clusters in this range with an approximately flat and uncorrelated distribution in the fraction of clusters recovered. 

A few dozen of their reported clusters may be genuinely challenging to detect in \emph{Gaia} data, since some of their clusters have $A_V > 5$ and are at high distances of greater than 10~kpc. However, the majority of their clusters are within 10~kpc and have $A_V < 5$. Given that \emph{Gaia} data have $\sim10^3$ times greater astrometric precision than \emph{Hipparcos} data for $\sim10^5$ times as many stars \citep{gaia_collaboration_gaia_2021}, and given that our chance of detecting a cluster reported in \cite{kharchenko_global_2013} is uncorrelated with extinction for $A_V < 5$, limitations of \emph{Gaia} data do not appear to be responsible for the bulk of non-detections of clusters from pre-\emph{Gaia} works, despite assertions in recent works that \emph{Gaia} data may be extinction-limited and unable to recover many highly reddened OCs from infrared datasets. Nevertheless, a handful of high-extinction clusters with $A_V > 5$ reported in the literature may still be challenging to recover in \emph{Gaia} data.

\subsection{The cluster does not exist}\label{sec:discussion-undetected:cluster-not-real}

Having exhausted all other major possibilities for why a cluster may not appear in our catalogue, the final potential reason would be that the cluster simply does not exist. As stated in the introduction to this section, far too many clusters are non-detected in this work for us to individually review them all and decisively prove that they are not real; however, we can give a broad overview of the typical characteristics of non-detected clusters, and contrast the similarities and differences between non-detected clusters in this work.


Figure~\ref{fig:crossmatch_comparison} shows that the parameter most strongly correlated with $f_\text{undetected}$ is the number of member stars $N$, with the smallest clusters from all papers being the least likely to be redetected. Few works report the statistical likelihood of a cluster being real in a way similar to the CST used in this work; however, $N$ can be thought of as a good proxy for the statistical significance of a cluster, as it stands that a cluster with fewer member stars is probably less likely to be real. Clusters with fewer than 20 reported sources are often the most difficult to redetect. 

In general, since most works in Fig.~\ref{fig:crossmatch_comparison} use \emph{Gaia} DR2 data or stronger cuts on \emph{Gaia} data than our methodology, there are many cases where we should be able to detect their reported clusters easily and with a higher number of member stars and statistical significance. The fact that we cannot suggests that some of these clusters may have been statistically insignificant associations of a small number of member stars.


The distance of undetected reported literature clusters is similarly revealing. In Sect.~\ref{sec:discussion-undetected:methodological-reasons:1-algorithm}, we suggest that some clusters may be undetected in this work at high distances due to limitations of the HDBSCAN algorithm. However, given that HDBSCAN should be the most sensitive algorithm for recovery of nearby clusters (Paper~1), it makes little sense that we are unable to recover a number of nearby clusters within $1~kpc$ for most of the works in Fig.~\ref{fig:crossmatch_comparison}. Many of these nearby and undetected objects may not be real, as there is no reason why we should not be able to detect them using the improved data of \emph{Gaia} DR3 and the most sensitive algorithm for recovery of nearby OCs.


The age of undetected clusters paints a complicated picture. In principle, detecting an old cluster has two challenges. Firstly, as the cluster ages, the brightest stars in the cluster evolve into faint remnants, which reduces the number of stars visible in the cluster. This is a particular issue for distant old clusters, as the remaining fainter and longer-lived stars in a cluster may be below a survey's magnitude limit. In the case of \emph{Gaia}, stars near to its magnitude limit have the lowest accuracy astrometry, reducing the signal-to-noise ratio of a given old, distant cluster in proper motion and parallax space -- further complicating its detection. Secondly, as clusters age, they are theorised to take a sparser and less centrally concentrated distribution \citep{portegies_zwart_young_2010}, reducing their signal-to-noise ratio relative to background field stars in positional data.

Although old clusters are likely to be harder to detect, in Paper~1, we found that the age of a reported cluster generally has the same effect on all algorithms: their lower number counts and sparsity affect all algorithms more or less equally in making them harder to detect. However, there are correlations between $f_\text{undetected}$ and $\log t$ for almost all papers in Fig.~\ref{fig:crossmatch_comparison}, despite all of them other than \cite{kharchenko_global_2013} being based on \emph{Gaia} data and using methods found in Paper~1 to be equally affected by cluster age. Hence, these correlations may be more informative about the types of cluster in other catalogues that are false positives than on whether or not a given catalogue used a better method.

For all works other than \cite{cantat-gaudin_clusters_2020}, clusters older than an age of around 1~Gyr ($\log t > 9$) are much less likely to be redetected. \cite{zucker_disconnecting_2022} have recently investigated the nature of the groups reported in \cite{kounkel_untangling_2020}, and find that many of them have ages $\sim$120 times larger than their dispersal times while being unbound and chemically homogeneous with their surrounding field stars -- strongly suggesting that they are merely associations of field stars and not physical groupings. The fact that we are unable to redetect almost any of the groups older than 1~Gyr reported in \cite{kounkel_untangling_2020} supports this conclusion, with it being plausible that many of their oldest groups are instead associations of field stars, consistent with the mean ages of field stars in the galactic thin and thick disks of a few Gyr. The similar correlations with old clusters being undetected for other works may also suggest that a number of other old clusters reported in the literature are also associations of field stars with mean ages similar to that of the typical ages of unclustered field stars in the galactic disk.

The reasons for the non-detection of some young clusters are less clear, and are more surprising given that young clusters should be easier to detect. In the case of \cite{cantat-gaudin_clusters_2020}, the handful of young clusters that we are unable to detect are also at high distances, which may mean that their non-detection is entirely a result of our own methodological limitations (see Sect.~\ref{sec:discussion-undetected:methodological-reasons:1-algorithm}.) On the other hand, these distant, young clusters may have originally been detected by hand-searching for OB stars in pre-\emph{Gaia} works and cataloguing them as OCs, but without a test of their physical nature, which could mean that they are associations. Similar reasoning could also be applied to the non-detected young clusters from \cite{kharchenko_global_2013}. Both possibilities are plausible, and this should be investigated further in another work.

\begin{figure*}[t]
   \centering
   \includegraphics[width=\textwidth]{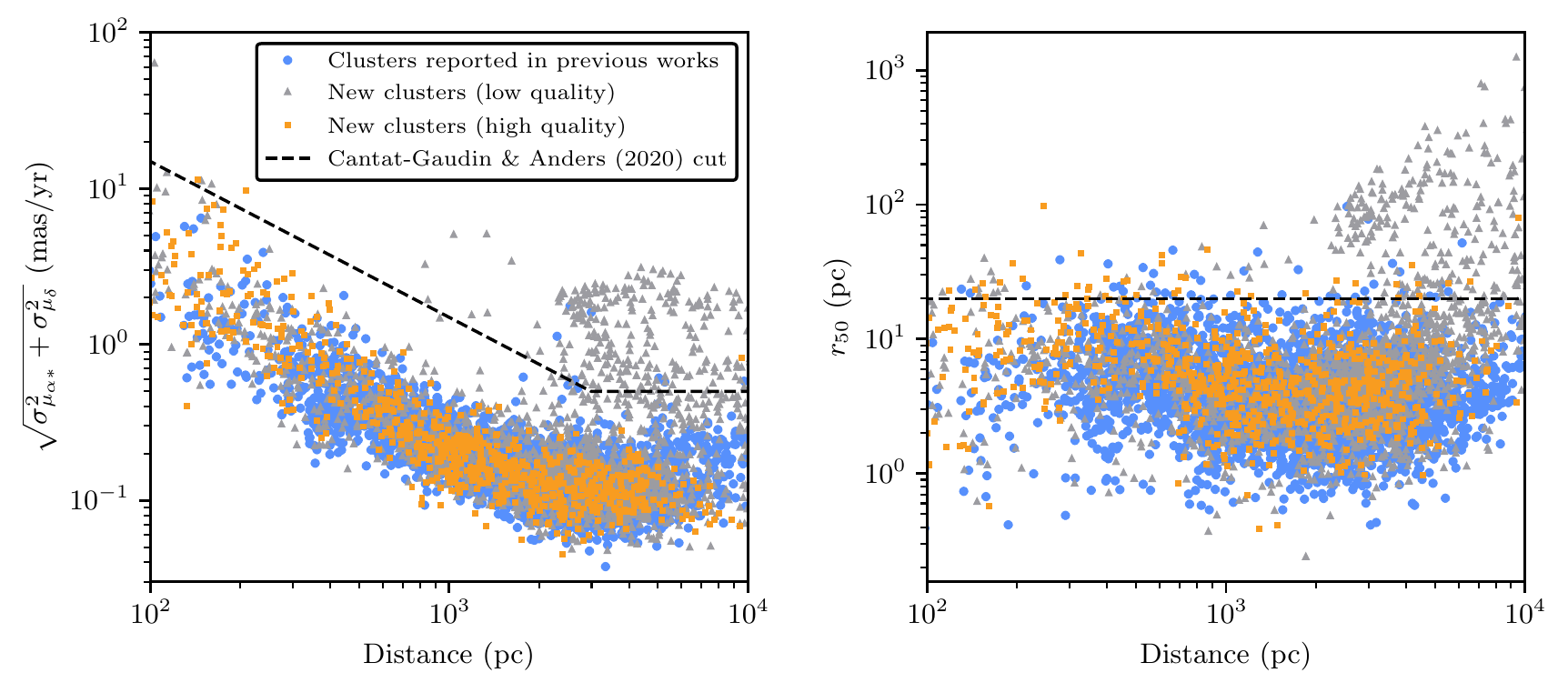}
   \caption{Geometric mean of the proper motion dispersion (left) and radius containing 50\% of members (right) for the clusters reported in this work, as a function of distance. Clusters are split between those detected in previous works (blue circles) and those newly reported in this work, divided between the high quality (orange squares) and low quality (grey triangles) samples defined in Sect.~\ref{sec:results-overall:suggested-cuts}. The cuts on cluster parameters to distinguish between bound OCs and unbound moving groups or associations proposed in \cite{cantat-gaudin_clusters_2020} are shown as a dashed black line.}%
   \label{fig:tcg_cut_comparison}
\end{figure*}

\begin{figure*}[t]
   \centering
   \includegraphics[width=\textwidth]{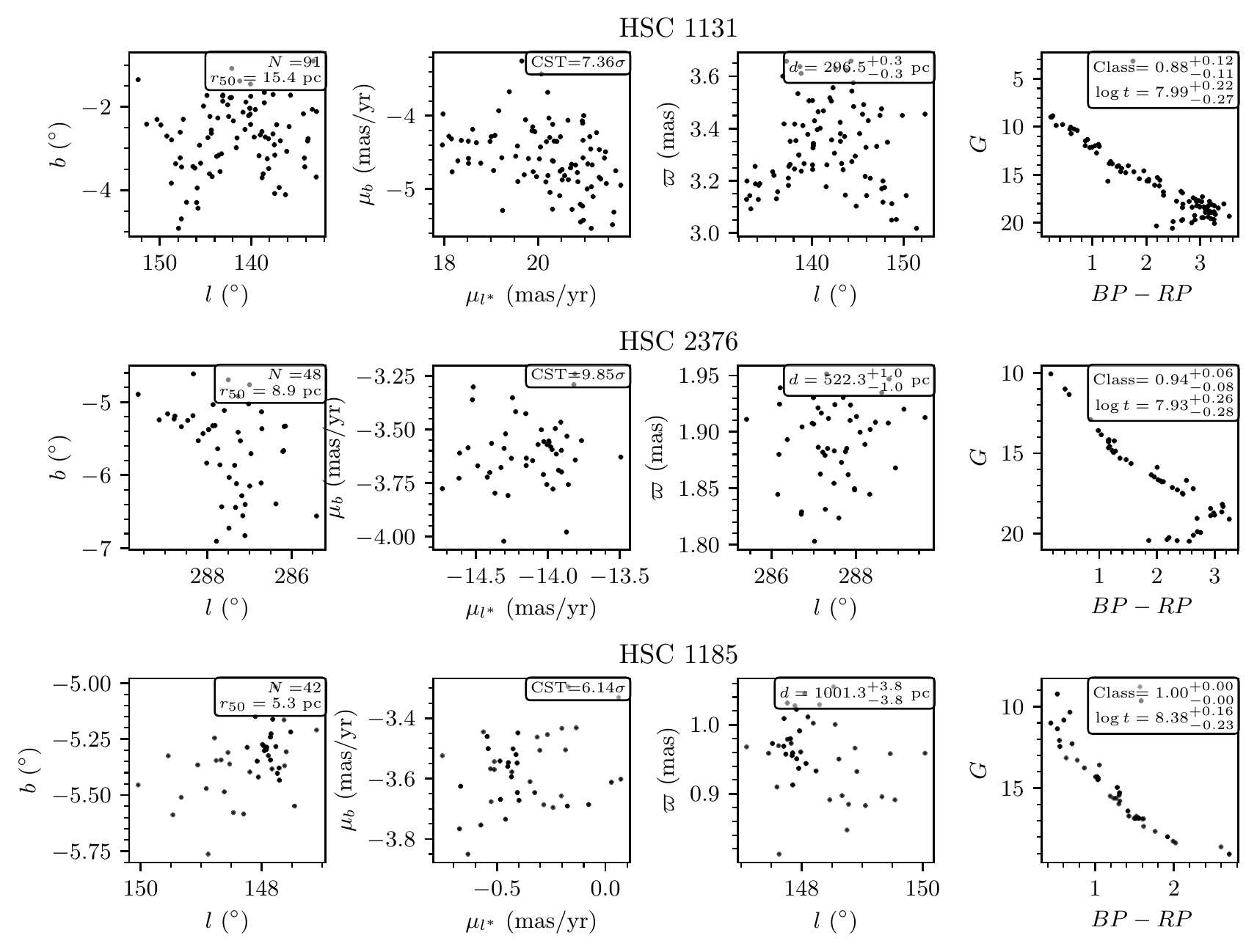}
   \caption{Three newly reported clusters randomly selected from the cluster catalogue and ordered by increasing distance, with member stars plotted as a function of their astrometric and photometric data as in Fig.~\ref{fig:membership_comparison}. All clusters pass the cuts proposed in \cite{cantat-gaudin_clusters_2020}, have good-quality CMDs passing the cuts from Sect.~\ref{sec:cmd_classifier}, and have astrometric significances of greater than 5$\sigma$, meaning they are almost certainly real overdensities in \emph{Gaia} data.}%
   \label{fig:sus_clusters}
\end{figure*}

Finally, the reasons for the spikes in non-detected clusters between $7 < \log t < 8$ for \cite{castro-ginard_hunting_2020, castro-ginard_hunting_2022} and between $6 < \log t < 7$ in \cite{hao_newly_2022} remain unclear. These works are entirely compiled from \emph{Gaia} DR2 and EDR3 data using the DBSCAN algorithm. Given that our results in Paper 1 suggest that clustering algorithms applied to \emph{Gaia} data have no differences between themselves in their ability to detect clusters based on their age, there is no clear reason why these clusters would be undetectable. The non-detection of these clusters should be investigated further.


For most works, extinction $A_V$ does not predict the chance of redetecting a given cluster. In Sect.~\ref{sec:discussion-undetected:methodological-reasons:5-gaia-limitations}, we discuss that $A_V$ values of greater than $\sim5$ appear to reduce the chance of a cluster being recovered in \emph{Gaia} data. The increasing trend in $f_\text{undetected}$ for \cite{cantat-gaudin_clusters_2020} as a function of $A_V$ appears to entirely be due to our lower chance of detecting clusters with $d>10$~kpc, since distant clusters also often have a high $A_V$. No other clear correlations exist for other works in Fig.~\ref{fig:crossmatch_comparison} with respect to extinction, other than for a few dozen pre-\emph{Gaia} clusters from the infra-red catalogue of \cite{kharchenko_global_2013} with $A_V \gtrsim 5$ that we are unable to redetect with \emph{Gaia} data.

In summary, we find that there are many potential reasons for the non-detection of given clusters from the literature, all of which should be investigated in more depth in future works. Verifying that new clusters reported in the literature are real is arguably as important as reporting them. While we cannot provide conclusive reasons for the non-detection of given clusters, given the scope of this survey, the overall trends we have identified should still be helpful and suggestive in whether or not given objects are real. We provide a table of all clusters non-detected by this work in Table~\ref{app:tab:all_crossmatches} and at the CDS.



\section{The difficulties of distinguishing between open clusters and moving groups}\label{sec:discussion-moving_groups}

Having discussed the catalogue's overall quality for the verification and study of clusters reported previously in the literature, it is worth discussing the 2387 new objects reported in this work -- 739 of which have a median CMD class above 0.5 and a CST of greater than $5\sigma$, and are hence the most reliable new objects that we report.

\subsection{The case against many of our new clusters being OCs}

On first inspection, despite having reliable CMDs and being statistically significant astrometric overdensities, many of our most reliable new objects have sparse density and proper motion distributions that appear more compatible with moving groups than spherically symmetric OCs with King \citep{king_structure_1962} or Plummer-like \citep{plummer_problem_1911} profiles. Figure~\ref{fig:sus_clusters} shows three clusters randomly selected from the 739 most reliable objects. HSC~1131 is a sparse, elongated grouping of stars in the thin disk, with a stringy nature much more compatible with a moving group than an OC. HSC~2376 is less clear, showing a more Gaussian clumping reminiscent of an OC within proper motion space but while still being relatively sparse, with $r_{50} = 8.9$~pc. HSC~1185 appears visually to be the most OC-like cluster, with its distribution of member stars forming compacter Gaussian-like overdensities in spatial and proper motion plots.

While we have used tests on statistical significance and cluster CMDs to determine the reliability of clusters in the catalogue, it is clear that a further test on the astrometric parameters of clusters (such as sparsity and proper motion dispersion) is necessary. \cite{cantat-gaudin_clusters_2020} propose two tolerant cuts on cluster parameters, finding that requiring the geometric mean of proper motion dispersion to be less than a criterion (corresponding to $\sim 5$~kms$^{-1}$) and $r_{50} < 20pc$ removed objects highly unlikely to be OCs from their sample.

However, Fig.~\ref{fig:tcg_cut_comparison} shows that with the exception of some clusters that are clearly associated with stellar streams (based on their location, CMD, and sparsity at distances greater than $\sim 3$~kpc), most new clusters detected in this work are compatible with OCs given the tolerant cuts in \cite{cantat-gaudin_clusters_2020}.

If almost all of the new clusters that we detect within 1~kpc of the Sun are in fact OCs, then this would represent a total paradigm shift in the census of OCs -- with a large number of previously unseen low number count, low mass, and sparse clusters being detectable nearby with \emph{Gaia} data. In reality, there are good reasons for this not being the case, and a more stringent cut on the astrometric parameters of candidate OCs is necessary.

In the preparation of this work, much effort was put in to attempting to find a more stringent cut on basic astrometric parameters (or some combination of them) to distinguish OCs from moving groups. We found that whether or not a cluster is a bound OC cannot be decided accurately based on individual cuts on $r_{50}$ or proper motion dispersions alone, and instead requires at least some modelling of the cluster's spatial profile, its velocity profile, and its mass. In the next section, we discuss the difficulties of such a method, which will be applied in the next paper in this series.

\subsection{A test for if our OC candidates are bound}

A given system is said to be in virial equilibrium if the absolute value of its potential energy $|V|$ is equal to twice its kinetic energy $T$. A number of works have recently used a relationship derived from the virial theorem, which predicts a velocity dispersion that a cluster should have if it is bound, $\sigma_\text{vir}$, based on its mass and radius. This can be compared to the cluster's measured 1D velocity dispersion $\sigma_{1D}$, which should equal $\sigma_\text{vir}$ if the cluster is bound:

\begin{equation}
    \sigma_\text{vir} = \sqrt{\frac{GM}{\eta r_\text{hm}}} \approx \sigma_{1D} \ \ \text{for a bound cluster,}
\end{equation}

\noindent
where $r_\text{hm}$ is the cluster's half-mass radius, $M$ is the cluster's mass, $G$ is the gravitational constant and $\eta$ is a constant depending on the cluster's density profile that is usually set to 10 \citep{portegies_zwart_young_2010}. In the case when $\sigma_{1D} >> \sigma_\text{vir}$, the cluster is likely to be unbound. This relationship has been used by works such as \cite{bravi_gaia-eso_2018}, \cite{kuhn_kinematics_2019}, and \cite{pang_3d_2021} to test the virial nature of OCs using \emph{Gaia} data, albeit in limited studies of no more than 28 clusters in one work.

While this relation is a promising way to distinguish between bound OCs and unbound moving groups, scaling this methodology to apply across our entire catalogue is extremely challenging. There are many systematics that can enter velocity dispersion, mass, and radius measurements, all of which must be reduced as much as possible to produce meaningful classifications. The clusters in our catalogue range across two orders of magnitude in distance, many orders of magnitude in mass, and two orders of magnitude in radius, with clusters of different parameters having fundamentally different challenges. For instance, nearby clusters may have tidal tails that must be removed from membership lists and may suffer from projection effects due to their radial velocity that would skew the measurement of their velocity dispersion with proper motions. On the other hand, distant clusters will push the limits of \emph{Gaia's} astrometric measurements, with velocity dispersions being difficult to measure precisely.

Given the scope of such a method, we leave its implementation to a future work. To restrict our catalogue to a reliable sample of OCs, users of our catalogue may for now use our CST scores, CMD classifications, and the criteria from \cite{cantat-gaudin_clusters_2020} to remove objects highly unlikely to be OCs. The next work to follow this one will provide a more accurate way to separate OCs from moving groups, and is anticipated to be submitted soon (Hunt \& Reffert, \emph{in prep.}).





\section{Conclusions and future prospects}\label{sec:conclusion}  

In this work, we conducted a blind all-sky search for Milky Way star clusters using \emph{Gaia} DR3 data. We show that a single blind search can be used to produce a homogeneous star cluster catalogue in the \emph{Gaia} era. We used the HDBSCAN algorithm, a density-based test of cluster significance, and a data partitioning scheme to detect as many reliable clusters as possible, producing a catalogue that is as complete and reliable as possible given current data. In total, the catalogue contains 7167 clusters, of which 4105 clusters form the most reliable sub-sample of objects with median CMD classifications greater than 0.5 and S/Ns greater than 5$\sigma$.

We provide a wide range of parameters for clusters in the catalogue, including: basic astrometric parameters, S/Ns that correspond to their statistical significance given \emph{Gaia} astrometry, CMD quality classifications, ages, extinctions, distances, and \emph{Gaia} DR3 radial velocities. We recover large, expansive membership lists for many OCs, often including tidal tails for clusters within $\sim 1$~kpc. Membership lists for all of our clusters are also available as a part of the catalogue (see Appendix~\ref{app:tables} and the CDS).

Extensive care was taken to crossmatch our catalogue against 35 other works. To the best of the authors' knowledge, these works catalogue all OCs reported in the literature, including many thousands of OCs recently reported in the literature using \emph{Gaia} data that are yet to be verified independently. 7022 clusters reported in the literature crossmatch against 4944 of the entries in our catalogue, including around 2000 of which we are able to independently verify for the first time. The spatial and age distribution of our catalogue traces the spiral arms in a similar way to many other recent works \citep[e.g.][]{cantat-gaudin_painting_2020, castro-ginard_milky_2021}.

However, we are unable to recover many of the clusters reported in the literature, despite our methodology having the highest sensitivity for OC recovery of all methods we trialed in Paper 1. We discuss reasons why we may be unable to detect an OC and are able to tentatively suggest that many thousands of clusters reported in the literature may not be real, including calling into question the common assertion that \emph{Gaia} is unable to recover a large fraction of OCs reported before \emph{Gaia} due to being extinction-limited. Further investigations into whether or not many of the OCs we are unable to detect are real would be helpful to improve the accuracy of the OC census.

Our catalogue contains 2387 new objects as yet unreported in the literature, 739 of which are a part of our most reliable sample of clusters with median CMD classifications of greater than 0.5 and an S/N of greater than $5\sigma$. While some of these objects are likely to be new OCs, we find that many are more compatible with unbound moving groups, as our methodology is sensitive to all kinds of stellar overdensity in \emph{Gaia} data. We find there is often no simple way to distinguish between the sparse, compact moving groups we detect and OCs, with the cuts on basic parameters proposed in \cite{cantat-gaudin_clusters_2020} being too lenient. In an upcoming work, we will use the virial theorem to distinguish between bound and unbound clusters with a probabilistic methodology (Hunt \& Reffert, \emph{in prep.}).

The coming decade of Milky Way star cluster research is likely to continue to be exciting and fast-paced. Firstly, the quality of available data will increase ever-higher. \emph{Gaia} DR4 will be produced from $\sim$66 months of data, almost double that of \emph{Gaia} DR3, which will result in a large jump in the accuracy of available astrometric and photometric data. DR4 is currently slated for release no sooner than the end of 2025. The current planned final \emph{Gaia} data release, DR5, may be based on around ten years of data, again roughly doubling the amount of input data used \citep{gaia_collaboration_gaia_2021}. Such large improvements in the accuracy of available astrometric data will inevitably result in more new clusters and improvements in the S/N and membership lists of existing clusters, further increasing the completeness and purity of the OC census.

Secondly, methodological improvements will continue to ease the process of star cluster recovery and characterisation. In the preparation of this work, it was still necessary to extensively verify many results by hand and develop postprocessing techniques to clean false positives from our catalogue. Improvements in clustering algorithms and techniques over the coming decade could make the process of cluster recovery more straightforward, accurate, and sensitive, with new methodologies such as Significance Mode Analysis (SigMA) methodology \citep{ratzenbock_significance_2022-1} showing promise in this area. As we discussed in Paper 1, there is currently no known perfect way to recover OCs from \emph{Gaia} data; much work remains to be done to try and find one.



\begin{acknowledgements}
We thank the anonymous referee for their helpful comments that improved the clarity of this work. We thank Tristan Cantat-Gaudin for reporting crossmatch issues to dwarf galaxies and NGC~6656 in an earlier version of this paper. E.L.H. and S.R. gratefully acknowledge funding by the Deutsche Forschungsgemeinschaft (DFG, German Research Foundation) -- Project-ID 138713538 -- SFB 881 (``The Milky Way System'', subproject B5). This work has made use of data from the European Space Agency (ESA) mission {\it Gaia} (\url{https://www.cosmos.esa.int/gaia}), processed by the {\it Gaia} Data Processing and Analysis Consortium (DPAC, \url{https://www.cosmos.esa.int/web/gaia/dpac/consortium}). Funding for the DPAC has been provided by national institutions, in particular the institutions participating in the {\it Gaia} Multilateral Agreement. This research has made use of NASA's Astrophysics Data System Bibliographic Services. This research also made use of the SIMBAD database, operated at CDS, Strasbourg, France \citep{wenger_m_simbad_2000}.

In addition to those cited in the main body of the text, this work made use of the open source Python packages NumPy \citep{harris_array_2020}, SciPy \citep{virtanen_scipy_2020}, IPython \citep{perez_ipython_2007}, Jupyter \citep{kluyver_jupyter_2016}, Matplotlib \citep{hunter_matplotlib_2007}, pandas \citep{mckinney_data_2010, jeff_reback_pandas-devpandas_2020}, Astropy \citep{robitaille_astropy_2013, astropy_collaboration_astropy_2018}, healpy \citep{zonca_healpy_2019}, and scikit-learn \cite{pedregosa_scikit-learn_2011}.

\end{acknowledgements}


\bibliographystyle{aa} 
\bibliography{references_abbreviated} 
%
%

\begin{appendix}

\section{Description of contents of online tables}\label{app:tables}

We provide tables of clusters, rejected clusters, member stars, and members stars for rejected clusters at the CDS. Tables of clusters follow the table format in Table~\ref{app:tab:cluster_lists}. Tables of members follow the same columns and column naming scheme as in \emph{Gaia} DR3 \citep{gaia_collaboration_gaia_2022}, except while also having columns referencing the cluster name and cluster ID we assign them to, the cluster membership probability, and a flag for if the star is a member within our estimated tidal radius $r_t$.

\begin{table}\label{app:tab:cluster_lists}
\caption{Description of the columns in the tables of detected clusters.}
\centering
\begin{tabular}{c c c l}
\hline\hline
Col. & Label & Unit & Description \\
\hline          
1     & Name                  & --  & Designation \\
2     & Internal ID           & --  & Internal designation \\
3     & All names             & --  & All literature names \\
4     & Kind                  & --  & Estimated object type\tablefootmark{c} \\
5     & $n_\text{stars}$      & --  & Num. of member stars \\
6     & $\text{S/N}$          & --  & Astrometric S/N \\
7     & $n_\text{stars}|_{r_\text{t}}$ & --  & $n_\text{stars}$ within $r_t$ \\
8     & $\text{S/N}|_{r_\text{t}}$ & --  & $\text{S/N}$ within $r_t$ \\
\hline

9-10     & $\alpha$, $\delta$ & deg & ICRS position \\
11-12    & $l$, $b$           & deg & Galactic position \\
13-16    & $r_{50,\,c,\,t,\,\text{tot}}$ & deg & Angular radii \\
17-20    & $R_{50,\,c,\,t,\,\text{tot}}$ & pc & Physical radii \\

21-26\tablefootmark{a} & $\mu_{\alpha^*}$, $\mu_\delta$ & mas yr$^{-1}$ & ICRS proper motions \\
27-29\tablefootmark{a} & $\varpi$              & mas & Parallax \\
30-32\tablefootmark{b} & $d$                   & pc  & Distance \\
33    & $n_d$                 & pc  & $n_\text{stars}$ for distance calc. \\
34    & $\varpi_0$ type       & --  & Parallax offset type\tablefootmark{d} \\
35-37 & $X$, $Y$, $Z$         & pc  & Galactocentric coords. \\

38-40\tablefootmark{a} & RV   & km s$^{-1}$ & Radial velocity\tablefootmark{e} \\
41    & $n_\text{RV}$         & --  & $n_\text{stars}$ with RVs \\
\hline

42-46\tablefootmark{b} & CMD class & --  & CMD class quantiles\tablefootmark{f} \\
47    & Human class           & --  & (where available)\tablefootmark{f} \\

48-50\tablefootmark{b} & $\log t$              & $\log \left[ \text{yr} \right]$  & Cluster age \\
51-53\tablefootmark{b} & $A_V$                 & mag & V-band extinction \\
54-56\tablefootmark{b} & $\Delta A_V$        & mag & Differential $A_V$ \\
57-59\tablefootmark{b} & $m-M$                 & mag & Photometric dist. mod. \\
\hline

60    & $m_{clSize}$          & --  & HDBSCAN parameter \\
61    & \texttt{merged}  & --  & Flag if merged\tablefootmark{g} \\
62    & \texttt{is\_gmm}  & --  & Flag if GMM used\tablefootmark{h} \\
63    & $n_\text{crossmatches}$ & --  & Num. crossmatches \\
64    & Xmatch type  & --  & Type of crossmatch\tablefootmark{i} \\

\hline
\end{tabular}

\tablefoot{The full version is available at the CDS.
\tablefoottext{a}{Mean value, standard deviation $\sigma$, and standard error $\sigma \, / \sqrt{n}$ are given.}
\tablefoottext{b}{Median value and various confidence intervals are given.}
\tablefoottext{c}{\texttt{g} for objects in the \cite{vasiliev_gaia_2021} GC catalogue, otherwise \texttt{o} (OC) or \texttt{m} (moving group) for clusters according to the empirical cuts in \cite{cantat-gaudin_clusters_2020}.}
\tablefoottext{d}{Flag indicating six clusters for which parallax bias correction using the method of \cite{lindegren_gaia_2021} was not possible, and a global offset was used instead (see Sect.~\ref{sec:clustering:parameters}).}
\tablefoottext{e}{Corrected using cluster distances to be relative to cluster centre.}
\tablefoottext{f}{Cluster CMD classes (the probability of a given cluster being a single coeval population of stars) derived using the neural network in Sect.~\ref{sec:cmd_classifier}. Some clusters also appeared in our human-labelled test dataset, for which human classes are also listed.}
\tablefoottext{g}{Indicates 25 clusters merged by hand where initial HDBSCAN clustering was unsatisfactory (see Sect.~\ref{sec:clustering}).}
\tablefoottext{h}{Indicates nine clusters with members from an additional Gaussian mixture model clustering step, typically applied to difficult to separate binary clusters.}
\tablefoottext{i}{Method used to assign name to cluster. In particular, `many to many' crossmatches are the most difficult to assign due to multiple objects crossmatching to multiple literature entries co-dependently (see Sect.~\ref{sec:crossmatching:names} for full discussion of final cluster name assignments.)}
}

\end{table}

\section{Table of crossmatch results}\label{app:crossmatches}

\begin{table*}[ht]
\caption{All cluster crossmatches, including literature clusters that have no match.\label{app:tab:all_crossmatches}}
\centering
\begin{tabular}{*{12}{c}}
\hline\hline
ID & Name & Source & Type & $\theta$ & $\theta_r$\tablefootmark{a} & $s_{\mu_{\alpha*}}$ & $\sigma_{\mu_{\alpha*}}$ & $s_{\mu_{\delta}}$ & $\sigma_{\mu_{\delta*}}$ & $s_{\varpi}$ & $\sigma_{\varpi}$ \\
& & & & ($^\circ$) & & (mas yr$^{-1}$) & & (mas yr$^{-1}$) & & (mas) \\
 
\hline

\multicolumn{12}{c}{$\cdot \cdot \cdot$} \\ 

176 & Basel 1 & Cantat-Gaudin+20 & gaia dr2 & 0.01 & 0.04 & 0.03 & 0.00 & 0.01 & 0.00 & 0.01 & 0.00\\
176 & Basel 1 & Dias+02 & position & 0.03 & 0.12 & - & - & - & - & - & -\\
176 & Basel 1 & Kharchenko+13 & hipparcos & 0.01 & 0.04 & 0.40 & 0.09 & 1.04 & 0.24 & 0.06 & 0.17\\
179 & Basel 10 & Bica+18 & position & 0.01 & 0.04 & - & - & - & - & - & -\\
179 & Basel 10 & Dias+02 & position & 0.01 & 0.04 & - & - & - & - & - & -\\
179 & Basel 10 & Cantat-Gaudin+20 & gaia dr2 & 0.01 & 0.07 & 0.03 & 0.00 & 0.05 & 0.01 & 0.01 & 0.00\\
179 & Basel 10 & Kharchenko+13 & hipparcos & 0.01 & 0.07 & 0.30 & 0.05 & 2.49 & 0.51 & 0.02 & 0.00\\
179 & Basel 10 & Kharchenko+13 & position & 0.01 & 0.07 & - & - & - & - & - & -\\
183 & Basel 11A & Cantat-Gaudin+20 & gaia dr2 & 0.01 & 0.01 & 0.02 & 0.00 & 0.05 & 0.00 & 0.02 & 0.00\\
183 & Basel 11A & Kharchenko+13 & hipparcos & 0.01 & 0.04 & 0.52 & 0.12 & 1.66 & 0.42 & 0.11 & 0.81\\
183 & Basel 11A & Dias+02 & position & 0.02 & 0.06 & - & - & - & - & - & -\\
183 & Basel 11A & Bica+18 & position & 0.03 & 0.06 & - & - & - & - & - & -\\
183 & Basel 11A & Kharchenko+13 & position & 0.01 & 0.04 & - & - & - & - & - & -\\
3003 & Basel 11B & Kharchenko+13 & position & 0.11 & 0.25 & - & - & - & - & - & -\\
184 & Basel 11B & Kharchenko+13 & hipparcos & 0.02 & 0.06 & 1.28 & 0.37 & 0.24 & 0.06 & 0.17 & 1.40\\
184 & Basel 11B & Kharchenko+13 & position & 0.02 & 0.06 & - & - & - & - & - & -\\
184 & Basel 11B & Dias+02 & position & 0.01 & 0.02 & - & - & - & - & - & -\\
184 & Basel 11B & Cantat-Gaudin+20 & gaia dr2 & 0.01 & 0.03 & 0.02 & 0.00 & 0.01 & 0.00 & 0.03 & 0.00\\
184 & Basel 11B & Bica+18 & position & 0.00 & 0.01 & - & - & - & - & - & -\\
6363 & Basel 11B & Kharchenko+13 & hipparcos & 0.11 & 0.39 & 2.15 & 0.64 & 1.99 & 0.59 & 0.22 & 1.98\\
6363 & Basel 11B & Kharchenko+13 & position & 0.11 & 0.39 & - & - & - & - & - & -\\
- & Basel 12 & Dias+02 & position & - & - & - & - & - & - & - & -\\
- & Basel 12 & Kharchenko+13 & hipparcos & - & - & - & - & - & - & - & -\\
- & Basel 12 & Bica+18 & position & - & - & - & - & - & - & - & -\\
180 & Basel 13 & Kharchenko+13 & position & 0.11 & 0.74 & - & - & - & - & - & -\\
- & Basel 13A & Bica+18 & position & - & - & - & - & - & - & - & -\\
- & Basel 14 & Dias+02 & position & - & - & - & - & - & - & - & -\\
- & Basel 14 & Kharchenko+13 & hipparcos & - & - & - & - & - & - & - & -\\
- & Basel 15 & Bica+18 & position & - & - & - & - & - & - & - & -\\
- & Basel 15 & Kharchenko+13 & hipparcos & - & - & - & - & - & - & - & -\\
- & Basel 15 & Dias+02 & position & - & - & - & - & - & - & - & -\\
181 & Basel 17 & Kharchenko+13 & position & 0.00 & 0.02 & - & - & - & - & - & -\\
181 & Basel 17 & Cantat-Gaudin+20 & gaia dr2 & 0.01 & 0.07 & 0.01 & 0.00 & 0.06 & 0.08 & 0.01 & 0.00\\
181 & Basel 17 & Kharchenko+13 & hipparcos & 0.00 & 0.02 & 0.62 & 0.10 & 2.37 & 0.43 & 0.11 & 0.93\\
181 & Basel 17 & Dias+02 & position & 0.00 & 0.02 & - & - & - & - & - & -\\

\multicolumn{12}{c}{$\cdot \cdot \cdot$} \\ 

\hline
\end{tabular}

\tablefoot{The full version is available at the CDS; the above only shows crossmatches against a selection of Basel clusters. Depending on the type of work crossmatched against, only separations in terms of position $\theta$ may be listed. For works with astrometry, separations $s$ with respect to $\mu_{\alpha*}$, $\mu_\delta$, and $\varpi$ are shown, in addition to separations $\sigma$ which are in terms of standard deviations about the mean of the astrometry of these clusters added together in quadrature, after accounting for worst-case systematics. Cluster entries in the literature that did not have a valid crossmatch against any cluster detected in this study are listed with only the name, source, and source type columns filled. Recalling Sect.~\ref{sec:crossmatching}, for a valid crossmatch, we require $\theta_r < 1$, and additionally, when crossmatching to a work with full five parameter astrometry, all $\sigma$ values to be less than two. 
\tablefoottext{a}{The separation between cluster centres in terms of the largest cluster radius available, $\theta_r = \theta / \text{max}(r_t, \, r_{t,\,\text{lit}})$}
}

\end{table*}

Here we provide a table of all crossmatches to all literature clusters that meet our adopted crossmatch criteria from Sect.~\ref{sec:crossmatching} in Table~\ref{app:tab:all_crossmatches}. For every cluster in the literature that we detect in this work, the table lists the internal cluster ID corresponding to our table of clusters in Table~\ref{tab:catalogue} that corresponds to this object. For clusters that we do not redetect, only a blank row with the cluster name, source paper, and type of crossmatch is shown.

\section{Bayesian neural networks}\label{app:bayesian_nets}

Given that Bayesian neural networks (BNNs) are only just beginning to see use in the astronomical literature \citep[e.g.][]{huertas-company_hubble_2019}, here we provide a brief background overview of the advantages and caveats of the approximate BNN methodology we adopted in Sect.~\ref{sec:cmd_classifier} and Sect.~\ref{sec:agenn}.

BNNs are a somewhat elusive area of open research in machine learning. Their appeal is clear: unlike a deterministic approach or an approach based on simply perturbing network inputs, a perfect BNN would be able to estimate both aleatoric uncertainties, which are uncertainties that result from random phenomena, such as uncertainty on photometric measurements; and epistemic uncertainties, which are uncertainties that result from a lack of knowledge about the underlying processes being modelled. For instance, any remaining gaps or issues in the simulated training data we use would cause a traditional deterministic neural network to always output an incorrect answer, whereas a probabilistic neural network should at least output a wide range of answers that demonstrate its uncertainty in such difficult cases \citep{goan_bayesian_2020, jospin_hands-bayesian_2022}.

In practice, there is currently no perfect BNN architecture, with all approaches having some flaws \citep{goan_bayesian_2020, jospin_hands-bayesian_2022}. While a Monte-Carlo Markov chain (MCMC)-based approach should in theory be superior, where every network weight has an arbitrary posterior distribution, MCMC-based BNNs are extremely difficult or impossible to train accurately, with current sampling techniques being inadequate \citep{goan_bayesian_2020}. In addition, BNNs are often time consuming to train. Instead, `variational inference' is widely used to approximate BNNs. In this technique, an ideal BNN is approximated by perturbing network features, approximating a BNN by `emphasising or de-emphasising' certain parts of a trained model when the model is sampled. This can then be used to estimate the epistemic uncertainty of a model by sampling a variational network multiple times.

Many approaches for variational inference exist in the literature, with a common approach being dropout regularisation as an approximation of a BNN \citep{gal_dropout_2015-1}, having also been used within astronomy \citep[e.g.][]{huertas-company_hubble_2019, leung_deep_2019}. However, this approximation is not inherently Bayesian \citep{hron_variational_2017}, and may be improved upon with recent developments in the literature. Another common approximation is to assume that all layer kernel and bias weights are drawn from simple distributions, such as independent Gaussian distributions. This allows for gradients during network training to be calculated straightforwardly using Bayes by backpropagation \citep{blundell_weight_2015}. This approximation can hold relatively well for (simple) neural networks, which often have normally distributed weights, but may cause underfitting on more complicated problems \citep{goan_bayesian_2020}. Due to the time-consuming nature of repeated samples of all kernel and bias posterior distributions, we also apply an approximation known as Flipout to more efficiently sample them with a lower runtime while preserving good training characteristics \citep{wen_flipout_2018}. Similar approaches using Bayes by backpropagation and Flipout have seen some use in the astronomy literature \citep[e.g.][]{lin_detection_2021}. We use the implementations of \texttt{DenseFlipout} and \texttt{Convolution2DFlipout} layers in TensorFlow Probability \citep{dillon_tensorflow_2017}, minimising the evidence lower bound (ELBO) loss \citep{blundell_weight_2015}.

In initial tests, these approximations produced network outputs with reliable uncertainty estimates that correspond well to the uncertainty inherent to classifying star cluster CMDs. It is worth noting from the literature that variational-inference based approaches are still more overconfident than a true BNN when applied to unseen data \citep{goan_bayesian_2020}, and that this approach is still an imperfect estimator of the true uncertainty of our model; nevertheless, our adopted method was found to be as accurate as a traditional deterministic network architecture of the same configuration when applied to our training data, but while providing an estimate of its uncertainty and without dramatically increasing runtime during training or sampling.

\end{appendix}

\end{document}